\documentclass[review,onefignum,onetabnum]{article}

\usepackage{latexsym}
\usepackage{amssymb,amsbsy,amsmath,amsfonts,amssymb,amscd,amsthm}
\usepackage{graphicx}
\usepackage{hyperref}
\usepackage{dsfont}
\usepackage{mathtools}
\usepackage{pdfsync}
\usepackage{epsfig,epstopdf}
\usepackage{caption}
\usepackage{xcolor}
\usepackage{hyperref}
\usepackage[title]{appendix}
\usepackage{verbatim}
\usepackage{bbm}
\usepackage{lipsum}
\usepackage{float}
\usepackage[font=small]{caption}
\usepackage{subcaption}
\usepackage{braket}
\usepackage{booktabs}

\newtheorem{rmrk}[]{Remark}

\DeclarePairedDelimiter {\abs}{\lvert}{\rvert}

\newcommand{\R}{\mathbb{R}}
\newcommand{\N}{\mathbb{N}}
\newcommand{\Z}{\mathbb{Z}}
\newcommand{\di}{\,\mathrm{d}}
\newcommand{\pt}{\partial_t}
\newcommand{\px}{\partial_x}
\newcommand{\pxx}{\partial_{xx}^2}
\newcommand{\pr}{\partial_r}

\newcommand{\beq}{\begin{equation}}
	\newcommand{\eeq}{\end{equation}}
\newcommand{\beqa}{\begin{eqnarray}}
	\newcommand{\eeqa}{\end{eqnarray}}

\newcommand{\rev}[1]{#1}

\renewcommand{\vec}{\boldsymbol}
\renewcommand{\epsilon}{\varepsilon}

\newcommand\bigO{
	\mathchoice
	{{\mathcal{O}}}
	{{\mathcal{O}}}
	{{\scriptstyle\mathcal{O}}}
	{\scalebox{.7}{$\scriptstyle\mathcal{O}$}}
}

\providecommand{\keywords}[1]
{
	\small	
	\textbf{\textit{Keywords---}} #1
}

\numberwithin{equation}{section}
\numberwithin{prpstn}{section}
\numberwithin{ass}{section}
\numberwithin{rmrk}{section}

\numberwithin{equation}{section}
\numberwithin{prpstn}{section}
\numberwithin{ass}{section}
\numberwithin{rmrk}{section}

\title{A hybrid discrete-continuum modelling approach for the interactions of the immune system with oncolytic viral infections\thanks{Corresponding author: David Morselli (d.morselli@ucl.ac.uk)\\
This research was partially supported by the Italian Ministry of Education, University and Research (MIUR) through the “Dipartimenti di Eccellenza” Programme (2018-2022) – Dipartimento di Scienze Matematiche “G. L. Lagrange”, Politecnico di Torino (CUP: E11G18000350001). MED and DM are members of GNFM (Gruppo Nazionale per la Fisica Matematica) of INdAM (Istituto Nazionale di Alta Matematica). ALJ acknowledges the Australian Research Council (ARC) Discovery Project (DP) DP230100025. FF acknowledges support from the Australian Research Council (ARC) via the Discovery Project (DP) DP230100485. We also acknowledge the support of the Australian National Health and Medical Research Council, through grant NHMRC IDEAS 2013058. Part of this work was performed on the OzSTAR national facility at Swinburne University of Technology. The OzSTAR program receives funding in part from the Astronomy National Collaborative Research Infrastructure Strategy (NCRIS) allocation provided by the Australian Government, and from the Victorian Higher Education State Investment Fund (VHESIF) provided by the Victorian Government.}}

\author{David Morselli,\thanks{Department of Mathematics, University College London, 25 Gordon Street, London WC1H 0AY, United Kingdom}  \thanks{Department of Mathematical Sciences ``G. L. Lagrange'', Politecnico di Torino, Corso Duca degli Abruzzi 24, 10129 Torino, Italy} \thanks{Department of Mathematics, School of Science, Computing and Engineering Technologies, Swinburne University of Technology, John St, 3122, Hawthorn, VIC, Australia}  \thanks{Department of Mathematics ``G. Peano'', Università di Torino, Via Carlo Alberto 10, 10124 Torino, Italy}
	\and
	Marcello Edoardo Delitala,\footnotemark[1]
	\and
	Adrianne L. Jenner,\thanks{School of Mathematical Sciences, Queensland University of Technology, George St, 4000, Brisbane, QLD, Australia}
	\and
	Federico Frascoli\footnotemark[2]
	}

\begin{document}
\maketitle
	
\begin{abstract}
Oncolytic virotherapy, which employs genetically engineered viruses to target cancer cells and stimulate anti-tumour immune response, has emerged as a promising therapeutic strategy. In our previous work, we developed a stochastic agent-based model elucidating the spatial dynamics of infected and uninfected cells within solid tumours. Building upon this foundation, we present a novel stochastic agent-based model to describe the intricate interplay between the virus and the immune system; the agents' dynamics are coupled with a balance equation for the concentration of the chemoattractant that guides the movement of immune cells. To better understand the macroscopic behavior, we derive a formal continuum limit of the model and compare it quantitatively to the individual-based simulations in two spatial dimensions. Furthermore, we describe the travelling waves of the three populations, with the uninfected proliferative cells trying to escape from the infected cells while immune cells infiltrate the tumour.

Simulations show a good agreement between agent-based approaches and numerical results for the continuum model. In certain parameter regimes, both the agent-based and continuum models exhibit oscillatory behavior, echoing Hopf bifurcations seen in non-spatial analogues. However, divergences between the models in specific cases highlight the critical role of stochasticity. Notably, we find that a premature immune response may undermine therapeutic efficacy, emphasising the importance of timing and modulation in combined immunovirotherapy approaches. This further suggests the importance of clinically improving the modulation of the immune response according to the tumour's characteristics and to the immune capabilities of the patients.
\end{abstract}	



\section{Introduction}
Oncolytic viruses are able to infect and kill cancer cells, while mostly sparing healthy tissues \cite{blanchette23,fountzilas17,Kelly2007651review,lawler17,russell18}. Despite their high potential as targeted cancer therapy, virotherapy is usually unable to eradicate a tumour alone; hence, most of the current efforts are devoted towards its combination with other therapies \cite{martin18}. One of the most promising of such combinations is with immunotherapies \cite{engeland22}, which has been tested in several clinical trials (such as Refs. \cite{andtbacka19,desjardins18}; we refer to Ref. \cite{engeland22} for a more comprehensive review). The ``avoidance of immune destruction'' is one of the hallmarks of cancer \cite{hanahan2011hallmarks}, therefore therapies that contribute to the activation of the immune system may play a central role in keeping a tumour under control and, if possible, in eradicating it. The interplay between oncolytic viruses and immune cells is twofold: oncolytic viruses are able to stimulate immune cells, not only against viral particles, but also against tumour cells; on the other hand, an immune response that targets the oncolytic virus may prevent an effective infection in the whole tumour, making virotherapy inefficient \cite{shi20}. The complexities of these dynamics motivate the use of mathematical models to gain a deeper understanding, with the goal of suggesting optimal treatment schedules for the combination of virotherapy and immunotherapies.

Several mathematical models have previously been adopted for the study of the interactions between oncolytic viruses and the immune system, including ordinary differential equations (ODEs) \cite{almuallem21,conte25,eftimie18,storey20,vithanage23,wodarz01}, partial differential equations (PDEs) \cite{friedman18,kim18,lee20,wu04}, stochastic agent-based models \cite{storey21,surendran23} and hybrid discrete-continuous multi-scale models \cite{jenner22}. While most of them restrict their attention to the systemic immune response, some others also explicitly model immunotherapies, such as immune checkpoint inhibitors \cite{friedman18,kim18,storey20,storey21,surendran23,vithanage23} and chimeric antigens receptor T-cells (CAR-T) \cite{conte25,mahasa22}. 

In general, individual-based models track individual cells, making it possible to represent processes happening at single cell-scale and easily include stochasticity; in this context, continuous fields are often used to model molecular elements, such as nutrients, leading to a \emph{hybrid} modelling approach. In contrast, deterministic continuum models describe volume fractions, hence the biological interpretation of the terms comprised in the model equations is less straight-forward and stochasticity cannot be included easily. On the other hand, this approach is particularly suitable to deal with large cell numbers for long time scales, as numerical simulations are faster than in the case of agents-based models and sometimes analytical results may be obtained. In order to combine the benefits of the two modelling approaches and gain a more comprehensive understanding of the biological system under study, a standard method is to derive a continuum macroscopic model from the underlying discrete or hybrid stochastic model (see, for example,  Refs. \cite{champagnat07,johnston15,lorenzimurray20, macfarlane22,penington11}; we refer to the introduction of Ref. \cite{chaplain20} for a more comprehensive literature review). The observation of significantly different behaviours of the two modelling instances would then suggest that stochasticity plays a key role in the phenomenon under investigation.

In our previous work \cite{morselli23}, we adopted this approach to model the infection of tumour cells due to oncolytic viruses in the absence of an immune response, taking into account two alternative sets of rules governing cell movement. Our results in the case of unrestricted cell movement show partial tumour remission for parameter values within the biologically meaningful range. The goal of the present work is to analyse the impact of the immune system in this situation, with the aim to determine whether eradication or long term control of the tumour are attainable, at least in the absence of relevant physical constraints.

Immune interactions with a tumour involve several different types of immune cells, which are stimulated and inhibited by a large number of molecules. An accurate description of these processes goes beyond the scope of the present work. In order to facilitate some theoretical understanding of the model, we restrict our attention to a single type of immune cell, namely cytotoxic T-cells, with the ability to kill both infected and uninfected tumour cells. We then assume that tumour cells secrete chemoattractant and immune cells follow the chemotactic stimuli towards the tumour (see Ref. \cite{painter19} and the references therein); this leads to a hybrid and multiscale modelling approach. Although the derivation of this kind of model from microscopic rules is well-known (see Ref. \cite{almeida22} for the specific case of immune interactions with cancer and Refs. \cite{bubba20,charteris14} for more general situations), we are not aware of any other work comparing agent-based and continuous models for the interactions between immune system and oncolytic viruses.

We consider a tumour with poor immune infiltration (i.e., a \emph{cold} tumour in the classification of Ref. \cite{galon19}) and assume that the infection by the oncolytic virus induces an immune anti-tumour response by increasing immune cell inflow and improving immune recognition; such a therapy combination is often defined immunovirotherapy \cite{engeland22}. We also assume that the immune killing rate can be enhanced (e.g., by inhibition of the PD-1 and PD-L1 checkpoints \cite{iwai02}) and we evaluate its consequences on the therapy. First, the spatially homogeneous ODE is considered, revealing that some parameter regions give rise to stable limit cycles: this is not surprising, as the same behaviour is also observed in similar models describing interactions of cancer with immune cells \cite{eftimie11review}, oncolytic virus  \cite{baabdulla23,jenner18biomath,jenner19,pooladvand21} and both together \cite{eftimie16}. Then, the effects of the oscillations are explored in the spatial models: in some situations we observe the extinction of infected agents even though the continuous model show recurrence of infection. Overall, our results suggest that the enhancement of the immune response may either increase or decrease the effectiveness of oncolytic virotherapy, depending on the time and location of the viral injection.

The article is organised as follows. In Section \ref{sec:model}, we introduce the agent-based model and present its continuum counterpart (a formal derivation is presented in Appendix \ref{app:derivation}). In Section \ref{sec:ode}, we study the equilibria of the spatially homogeneous ODE and the emergence of a stable limit cycle; this analysis provides some insights on the oscillations observed in the full system. In Section \ref{sec:results}, we compare the results of numerical simulations of the agent-based model and the numerical solutions of the corresponding PDEs, comparing it with the situation in which the immune response is negligible. In Section \ref{sec:conclusions}, we discuss the main findings and provide some suggestions for future research.

\section{Description of the agent-based model and formal derivation of the corresponding continuum model}
\label{sec:model}

In our previous work \cite{morselli23}, we presented a stochastic agent-based model describing infected and uninfected cells for solid tumours that interact with viruses in the absence of an immune response. Our model takes into account proliferation and death of uninfected tumour cells, death of infected tumour cells, infection of uninfected cells and cell movement. We considered two alternative sets of rules governing the latter process (namely, undirected random cell movement and pressure-driven cell movement) and we showed how this choice strongly influences therapy outcomes. 

In this approach, we did not explicitly take into account viral dynamics: we here briefly motivate this choice, as we adopt the same strategy in the present work. While it is known that oncolytic viruses are able to infect through specific receptors that are highly expressed on cancer cells \cite{lawler17}, the exact mechanisms of the infection are not well understood. In general, infections may take place both through direct cell-to-cell transmission and cell-free transmission mediated by diffusing virions; the actual combination of the two processes is hard to establish in full detail (see Ref. \cite{graw16} and the references therein). In the specific case of oncolytic virotherapy, there are several obstacles to viral diffusion in the tumour microenvironment: indeed, factors such as extracellular matrix composition, immune cell infiltration and hypoxic regions can impede viral penetration, replication and spread within the tumour \cite{wojton10}. It is therefore a reasonable approximation to neglect viral spread far from infected cells and assume that the infection is mainly driven by cell-to-cell contact and close-range free virions. This approach has been commonly used for nonspatial models of oncolytic viruses \cite{komarova10,novozhilov06}. In the context of spatial models, this choice has also been motivated by the above-mentioned obstacles of viral diffusion in the tumour microenvironment

Our results suggest that the inability of free virions to propagate in the tumour microenvironment combined with constraints of cellular movement may cause the failure of the therapy due to stochastic effects. Therefore, in the present work we only focus on undirected movement, which is more likely to result in more favourable outcomes.

The details of our previous agent-based model are summarised in the context of the new model for the sake of completeness. The corresponding continuum model is the following diffusive Lotka--Volterra model with logistic growth:
\begin{equation}
	\label{eq:l}
	\begin{cases}
		\pt u(t,x)=D\pxx u(t,x)+pu(t,x)\Bigl(1-\dfrac{u(t,x)+i(t,x)}{K}\Bigr)-\dfrac{\beta}{K} u(t,x)i(t,x), \\[8pt]
		\pt i(t,x)=D\pxx i(t,x)+\dfrac{\beta}{K} u(t,x)i(t,x)-qi(t,x).
	\end{cases}
\end{equation}

We summarise in \ref{subsec:bmb} the main takeaways from this approach, which help to elucidate some aspects of the current work. The rest of the current section is devoted to presenting an extension of these discrete and continuous models.

\subsection{Agent-based model}

We extend and improve upon the modelling framework mentioned above by including immune cells, which are described as agents that occupy a position on a discrete lattice in the same way as cancer cells. We also consider a chemoattractant secreted by cancer cells that guide the movement of immune cells; its concentration is described as a discrete, non-negative function. Observe that we are using a hybrid discrete-continuous modelling framework, since the chemoattractant concentration is the discretisation of a continuous function. For ease of presentation, in this section, we restrict our attention to one spatial dimension, but there would be no additional difficulty in considering higher spatial dimensions. In the following sections, we mainly focus on two spatial dimensions, so, in Remark \ref{rmrk:twod}, we explain the small difference in this situation.

\begin{figure}
	\centering
	\includegraphics[width=\textwidth]{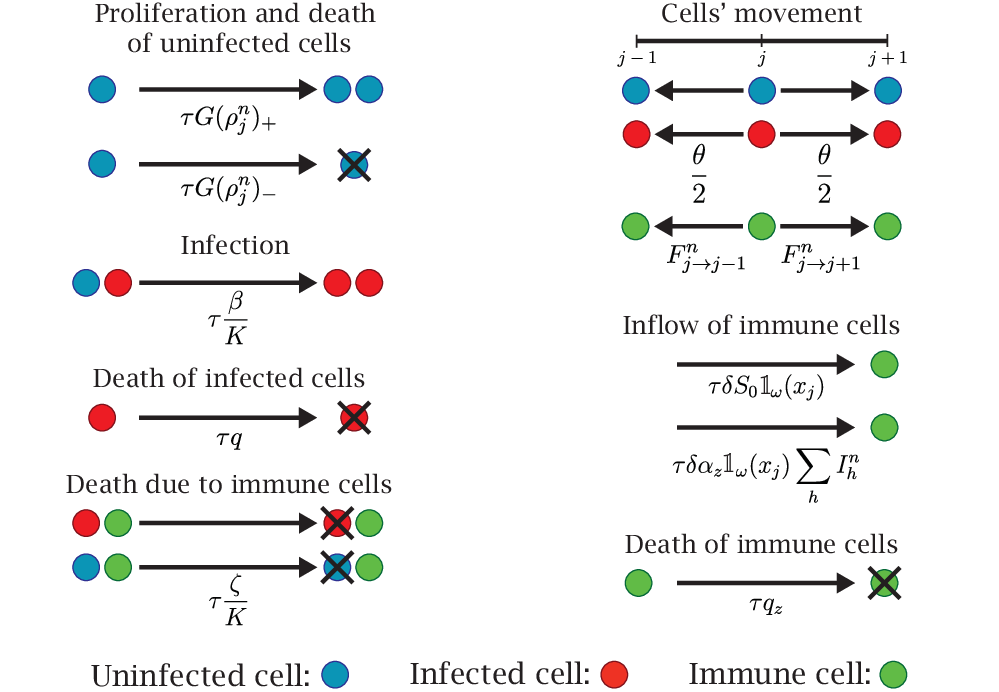}
	\caption{Schematic representation of the rules governing cell dynamics in the stochastic models. Uninfected cells are represented in blue, infected cells in red and immune cells in green. Uninfected cells may proliferate or die according to the total density, move, become infected upon contact with infected cells and die upon contact with immune cells. Infected cells may move, die with constant probability and die upon contact with immune cells. Immune cells may enter the domain, move with the probabilities given in Eq. \eqref{eq:F} and die with constant probability. \rev{The model also considers the dynamics of the chemoattractant, which are not included in the figure due to the different modelling approach adopted (i.e., density-based and deterministic instead of individual-based and stochastic).}}
	\label{fig:discrete}
\end{figure}

Let us consider the temporal discretisation $t_n=\tau n$, with $n\in\N_0$, $0<\tau \ll 1$, and the spatial domain $\Omega\subseteq\R$ with discretisation $x_j=\delta j$, with $j\in\Z$, $0<\delta\ll 1$; we assume $\tau$ to be small enough to guarantee that all the probabilities defined hereafter are smaller than $1$. We denote the number of immune cells, uninfected and infected cancer cells that occupy position $x_j$ at time $t_n$ respectively by $Z_j^n$, $U_j^n$ and $I_j^n$; the corresponding densities are
\[
z_j^n\coloneqq \frac{Z_j^n}{\delta}, \qquad u_j^n\coloneqq \frac{U_j^n}{\delta}, \qquad i_j^n\coloneqq \frac{I_j^n}{\delta}. 
\]
We then denote by $\phi_j^n$ the concentration of chemoattractant at time $t_n$ and position $x_j$. Since the spatio-temporal scales for the chemoattractant's dynamics are very different from cellular ones, we describe them with a deterministic discrete balance equation, as in Refs. \cite{almeida22,cooper14}. Table \ref{tab:variables} summarises the variables of the hybrid agent-based model and their macroscopic counterparts; Fig. \ref{fig:discrete} summarises the rules governing the dynamics of the agent-based model. Cancer cells proliferate, move, become infected and die in the same way as in Ref. \cite{morselli23}. The dynamics of the chemoattractant and of the immune cells represent a novelty with respect to our previous work and resemble some other models in the literature, as explained in the following. We assume that the infection stimulates the immune system by increasing the number of immune cells in the area and guiding them towards infected cells; once an immune cell comes in contact with a cancer cell, it is able to kill it even if it is not infected.

In what follows, for the sake of clarity, parameters related to similar processes are labelled in a consistent way (e.g., $q$, $q_z$, and $q_\phi$ for decay rates, $\alpha_\phi$ and $\alpha_z$ for secretion/inflow rates).

\begin{table}[t!]
	\centering{
		\scriptsize
		\begin{tabular}{llcc}
			&\multicolumn{1}{c}{\textbf{Quantity}} & \multicolumn{1}{c}{\textbf{Microscopic variable [Units]}} &  \multicolumn{1}{c}{\textbf{Macroscopic variable [Units]}} \\
			\hline
			\\
			& uninfected cancer cells         & $U_j^n$ [cells]     & $u(t,x)$ [cells/mm$^2$]\\
			& infected cancer cells         & $I_j^n$ [cells]     & $i(t,x)$ [cells/mm$^2$]\\
			& immune cells         & $Z_j^n$ [cells]    & $z(t,x)$ [cells/mm$^2$]\\
			& chemoattractant        & $\phi_j^n$ [$\mu$g/mm$^2$]    & $\phi(t,x)$ [$\mu$g/mm$^2$]\\
			\\
			\hline
		\end{tabular}
		\caption{List of the variables for both approaches, with their units of measurement.}\label{tab:variables}}
\end{table}

\paragraph{Basic dynamics of uninfected cancer cells}
We let an uninfected cell that occupies position $x_j$ at time $t_n$ reproduce with probability $\tau G(\rho_j^n)_+$, die with probability $\tau G(\rho_j^n)_-$, and remain quiescent with probability $1-\tau G(\rho_j^n)_+-\tau G(\rho_j^n)_-=1-\tau \abs{G(\rho_j^n)}$. Here $\rho_j^n\coloneqq u_j^n+i_j^n$ is the total cell density at time $t_n$ and position $x_j$ and
\begin{equation}
	\label{eq:G}
	G(\rho)=p\Bigl(1-\frac{\rho}{K}\,\Bigr),
\end{equation}
where $p>0$ is the maximal duplication rate and $K>0$ is the carrying capacity.

We consider undirected random movement and assume that an uninfected cell moves to an adjacent lattice point with probability $\theta/2$, where $\theta\in[0,1]$, and remains at its initial position with probability $1-\theta$.

We do not model explicitly the dynamics of the oncolytic virus, as previously explained; we instead assume that an uninfected cell that occupies position $x_j$ at time $t_n$ becomes infected upon contact with infected cells with probability $\tau\beta i_j^n/K$, where $K$ is the carrying capacity and $\beta>0$ is a constant infection rate. Let us remark that the parameter $\beta$ summarises different biological processes and, as a consequence, its estimate takes into account the death rate $q$ and other parameters related to viral dynamics (we refer to Ref. \cite{morselli26viral} for further discussions).
	
Our approach does not consider virus clearance due to the immune system. We remark that this process is not well-understood either and it is indeed neglected in several existing models \cite{conte25,eftimie11multi,eftimie16}. The discussion of the next sections shows that the immune response against infected cells in general decreases the effectiveness of viral infection, in a way similar to viral clearance. It is therefore interesting to investigate whether tumour eradication through immunovirotherapy could be achieved under ideal conditions.

\paragraph{Basic dynamics of infected cancer cells}
We assume that the infection does not affect the cell motility and so the probabilities are the same as the uninfected cells. We also assume that at every time step an infected cell may die because of lysis with probability $\tau q$, where $q>0$ is a constant death rate. We assume that the viral replication process hijacks the cells proliferation machinery and hence infected cells are unable to proliferate.

\paragraph{Dynamics of the chemoattractant}
We assume that uninfected and infected cells produce chemoattractant at rates $\gamma_\phi$ and $\alpha_\phi$, respectively. We choose their values so that $\alpha_\phi\gg\gamma_\phi>0$, in line with our assumption that the tumour is initially \emph{cold} and the infection by the oncolytic virus is enough to induce an anti-tumour immune response, as often observed \textit{in vivo} and \textit{in vitro} \cite{galon19}. Chemoattractant density is bounded and saturates at $\phi^*>0$. The chemoattractant also decays at rate $q_\phi>0$ and diffuses. The resulting balance equation is
\begin{equation}
	\label{eq:phi_discrete}
	\phi_j^{n+1}=\phi_j^n+\tau D_\phi \frac{\phi_{j+1}^n+\phi_{j-1}^n-2\phi_j^n}{\delta^2} +\tau (\alpha_\phi i_j^n+\gamma_\phi u_j^n) \rev{(\phi^*-\phi_j^n)}-\tau q_\phi \phi_j^n,
\end{equation}
where $D_\phi>0$ is the diffusion coefficient and $q_\phi$ the decay rate. This equation closely resembles the ones used in Refs. \cite{almeida22,bubba20} to model the evolution of a chemoattractant concentration.

\paragraph{Dynamics of immune cells}
We assume that there is a constant influx of immune cells into the microenvironment independent of the presence of cancer cells. In addition to this, we assume that infection by the oncolytic virus stimulates an immune response in the whole tumour. Hence, an immune cell appears at point $x_j$ at time step $t_n$ with probability $\tau \delta S_j^n$, given by
\begin{equation}
	\label{eq:S}
	S_j^n=\biggl(S_0+\alpha_z  \sum_h I_h^n \biggr)\mathbbm{1}_\omega (x_j),
\end{equation}
where $\mathbbm{1}_\omega$ is the indicator function of the set $\omega\subset \Omega$, $S_0>0$ is the base inflow rate and $\alpha_z>0$ is the additional inflow rate due to the infection\rev{; the latter takes into account the total number of infected cells in the domain}. In principle, we could vary $\omega$ to model the fact that some areas of the tumour are harder to reach for immune cells (e.g. due to poor vascularisation), although this goes beyond the scope of the present work. It is important to observe that the increase of the inflow due to infected cells is nonlocal, as in Ref. \cite{almeida22}; this resembles the recruitment of immune cells from adjacent lymph nodes and the subsequent arrival through blood vessels.

We then assume that an immune cell that occupies position $x_j$ at time $t_n$ moves to the lattice point $x_{j\pm 1}$ with probability $F_{j\to j\pm 1}^n$ and remains at its initial position with probability $1-F_{j\to j- 1}^n-F_{j\to j+ 1}^n$. We consider both undirected random movement and chemotaxis up the chemoattractant gradient, where the latter depends on the concentration difference between the cell’s initial position and the target point. We therefore set
\begin{equation}
	\label{eq:F}
	F_{j\to j\pm 1}^n\coloneqq\frac{\theta_z}{2} +\nu\frac{(\phi_{j\pm 1}^n-\phi_j^n)_+}{2\phi^*},
\end{equation}
where $z_+\coloneqq\max\{z,0\}$, $\phi^*$ is the saturation density of the chemoattractant and $\theta_z, \nu\in[0,1]$ with $\theta_z+\nu<1$. Observe that, if $0\leq \phi_j^n\leq \phi^*$ for every $j$, then all the probabilities are between $0$ and $1$. This kind of reasoning and the probabilities associated have already been employed in Refs. \cite{almeida22,bubba20}. While this modelling approach can include obstacles to immune infiltration, we neglect this aspect for simplicity. Let us remark that long-range viral spread is hindered not only by the extracellular matrix and other physical obstacles, but also by other processes such as viral clearance \cite{morselli26viral}; on the other hand, immune cells are not necessarily affected in the same way, as they rely on active, directed motility rather than passive diffusion. It is therefore reasonable to consider obstacles for viral diffusion and not for immune cell movement. 

Finally, we assume that at every time step an immune cell dies with probability $\tau q_z$, where $q_z>0$ is a constant death rate.

\paragraph{Cytotoxic action of the immune cells}
We assume that cancer cells may be killed by the cytotoxic action of immune cells upon contact; this happens at a rate proportional to the density of immune cells. To be precise, a cancer cell that occupies position $x_j$ at time $t_n$ dies with probability $\tau \zeta z_j^n/K$, where $K$ is the carrying capacity and $\zeta>0$ is a constant killing rate. For simplicity, we assume that the killing rate is the same for every cancer cell, although it could make sense to consider situations in which infected cells are more easily recognised by immune cells and, thus, are killed at a higher rate. This process is analogous to the infection of cancer cells described above.

\subsection{Corresponding continuum model}

Letting $\tau,\delta\to 0$ in such a way that $\frac{\delta^2}{2\tau}\to \tilde{D}$ and assuming that there are the functions $u\in C^2([0,+\infty)\times\R)$ such that $u_{j}^{n}=u(t_n,x_j)$, $i\in C^2([0,+\infty)\times\R)$ such that $i_{j}^{n}=i(t_n,x_j)$, $z\in C^2([0,+\infty)\times\R)$ such that $z_{j}^{n}=z(t_n,x_j)$ and $\phi\in C^2([0,+\infty)\times\R)$ such that $\phi_{j}^{n}=\phi(t_n,x_j)$ we formally obtain (see Appendix \ref{app:derivation}) the following system of reaction-diffusion PDEs
\begin{equation}
	\label{eq:pde}
	\begin{cases}
		\pt u(t,x)=D\pxx u(t,x)+pu(t,x)\Bigl(1-\dfrac{u(t,x)+i(t,x)}{K}\Bigr)-\dfrac{\beta}{K} u(t,x)i(t,x) \\[8pt]
		\phantom{\pt u(t,x)}-\dfrac{\zeta}{K} u(t,x)z(t,x), \\[8pt]
		\pt i(t,x)=D\pxx i(t,x)+\dfrac{\beta}{K} u(t,x)i(t,x)-qi(t,x)-\dfrac{\zeta}{K} i(t,x)z(t,x), \\[8pt]
		\pt z(t,x)=D_z\pxx z(t,x)-\dfrac{\chi}{\phi_\text{max}} \px(z(t,x) \px \phi(t,x)) -q_z z(t,x)+ S(t,x), \\[8pt]
		\pt \phi(t,x)=D_\phi \pxx \phi(t,x)+(\alpha_\phi i(t,x) +\gamma_\phi u(t,x))\,(\phi(t,x)-\phi^*)-q_\phi \phi(t,x).
	\end{cases}
\end{equation}
where $D\coloneqq \theta \tilde{D}$, $D_z\coloneqq \theta_z \tilde{D}$, $\chi\coloneqq \nu \tilde{D}$ and
\[
S(t,x)\coloneqq \biggl( S_0+ \alpha_z \int_\Omega i(t,y) \di y \biggr) \mathbbm{1}_\omega (x).
\]
The first two equations are the ones of Eq. \eqref{eq:l} with the addition of the death term related to the immune system; we therefore expect to recover similar results for small $\zeta$. This system resembles some of the models discussed in Ref. \cite{painter19} for the interactions between cancer and different kinds of immune cells, with the relevant differences being that one of our equations is integro-differential (as in Ref. \cite{almeida22}) and that the infection significantly affects the dynamics, spatially and temporally.

In the next Section we consider the two-dimensional radially equivalent version of this problem. Hence, we assume that
\begin{equation}
	\label{eq:omega}
	\omega\coloneqq \Set{\vec{x}\in \Omega \ \Big| \ \abs{\vec{x}}\leq R},
\end{equation}
with $R>0$; this corresponds to the situation of a well-vascularised tumour in which immune cells can easily reach any point of the domain or that of a solid tumour that is easily accessible by the immune system both from the histological and topological point of view. The system of PDEs then becomes
\begin{equation}
	\label{eq:pderad}
	\begin{cases} 
		\pt u=D\dfrac{1}{r} \pr (r\,\pr u)+pu\Bigl(1-\dfrac{u+i}{K}\Bigr)-\dfrac{\beta}{K} ui-\dfrac{\zeta}{K} uz, \\[8pt]
		\pt i=D\dfrac{1}{r} \pr (r\,\pr i)+\dfrac{\beta}{K} ui-qi-\dfrac{\zeta}{K} iz, \\[8pt]
		\pt z=D_z\dfrac{1}{r} \pr (r\,\pr z)-\dfrac{\chi}{\phi^*} \dfrac{1}{r} \pr (r z \, \pr \phi) -q_z z+ S, \\[8pt]
		\pt \phi=D_\phi \dfrac{1}{r} \pr (r\,\pr \phi)+(\alpha_\phi i +\gamma_\phi u)\,(\phi^*-\phi)-q_\phi \phi.
	\end{cases}
\end{equation}
with 
\begin{equation}
\label{eq:S_cont}
S(t,r)\coloneqq \biggl( S_0+ 2\pi \alpha_z \int_0^R i(t,s) s \di s \biggr) \mathbbm{1}_{[0,R]} (r).
\end{equation}

\begin{rmrk}
	\label{rmrk:twod}
	When the spatial domain is the two-dimensional real plane $\R^2$ instead of the one-dimensional real line $\R$, the scalar index $j\in\Z$ should be replaced by the vector $\vec{j}=(j_x,j_y)\in\Z^2$ and the probability that a cell moves to one of the four neighbouring lattice points is $\theta_k/4$, with $k=u,i$. We then need to scale $\tau$ and $\delta$ in such a way that $\frac{\delta^2}{4\tau}\to \tilde{D}$.
\end{rmrk}

\section{Corresponding ODE model and bifurcation analysis}
\label{sec:ode}

\begin{table}[t!]
	\centering{
		\scriptsize
		\begin{tabular}{llp{0.3\linewidth}lp{0.2\linewidth}}
			&\multicolumn{1}{c}{\textbf{Parameter}} & \multicolumn{1}{c}{\textbf{Description}} &  \multicolumn{1}{c}{\textbf{Value [Units]}} & \multicolumn{1}{c}{\textbf{Reference}}\\
			\toprule
			\\
			& $D$         &diffusion coefficients of cancer cells    & $1.88\times 10^{-4}$ [mm$^2$/h]	&estimate based on \cite{kim06} \\
			& $p$         &maximal duplication rate of uninfected cells     & $1.87 \times 10^{-2}$ [h$^{-1}$] &\cite{ke00} \\
			& $K$         &tissue carrying capacity in two dimensions   & $10^4$ [cells/mm$^2$]	& \cite{lodish08} \\
			& $\beta$         &infection rate   & $1.02\times 10^{-1}$ [h$^{-1}$]	&estimate based on \cite{friedman06} \\
			& $\zeta$         &immune killing rate of cancer cells   & $0.50$ or $5.00$ [h$^{-1}$]	&model estimate \\
			& $q$         &death rate of infected cells     & $8.34\times 10^{-3}$ [h$^{-1}$]	& model estimate \\	
			\midrule
			& $D_z$         &diffusion coefficients of immune cells   & $4.20\times 10^{-3}$ [mm$^2$/h]	& \cite{almeida22} \\
			& $\chi$         &chemotactic coefficient of immune cells   & $1.65\times 10^{-1}$ [mm$^2$/h]		&model estimate\\
			& $\phi^*$         &saturation density of chemoattractant   & $2.92$ [$\mu$g/mm$^2$]	& \cite{gao14} \\
			& $q_z$         &death rate of immune cells   & $7.50\times 10^{-3}$ [h$^{-1}$]	&\cite{hao2014mathematical} \\
			& $S_0$         &base inflow rate of immune cells  & $5.00\times 10^{-2}$ [cells/(mm$^2\cdot$h)]	&model estimate \\
			& $\alpha_z$         &inflow rate of immune cells due to the infection   & $3.75\times 10^{-5}$ [(mm$^2\cdot$h)$^{-1}$]	&model estimate \\
			\midrule
			& $D_\phi$         &diffusion coefficients of chemoattractant   & $3.33\times 10^{-2}$ [mm$^2$/h]	&\cite{matzavinos04} \\
			& $\alpha_\phi$         &secretion of chemoattractant by infected cells   & $2.50\times 10^{-4}$ [mm$^2$/(h$\cdot$cells)]	&model estimate \\
			& $\gamma_\phi$         &secretion of chemoattractant by uninfected cells   & $5.00\times 10^{-6}$ [mm$^2$/(h$\cdot$cells)]	&model estimate \\
			& $q_\phi$         &decay of chemoattractant   & $8.33\times 10^{-2}$ [h$^{-1}$]	&\cite{cooper14} \\
			\midrule
			& $R_u$         &initial radius of uninfected cells     & $2.60$ [mm]	&\cite{kim06} \\
			& $R_i$         &initial radius of infected cells     & $1.00$ [mm]	&model estimate \\ 
			\midrule
			& $\alpha$         &inflow rate of immune cells due to the infection (ODE)   & $2.94\times 10^{-3}$ [h$^{-1}$]	&model estimate \\
			\bottomrule
		\end{tabular}
		\caption{Reference parameter set used in this work.}
		\label{tab:parameters}}
\end{table}

Before comparing the agent-based and the continuous model, it is useful to consider a homogeneous spatial configuration and analyse the equilibria of the corresponding ODE model and their stability. The chemoattractant has the sole purpose of guiding immune cells, therefore it can be neglected in this nonspatial model. Let us also remark that the nonlocal activation of the immune system can only be modelled accurately by assuming that $u$, $i$ and $z$ are homogeneous in the spatial domain $\Omega$ and $\omega=\Omega$: under this assumption, the integral of Eq.~\eqref{eq:S_cont} becomes
\[
\mathbbm{1}_\omega (x)\int_\Omega i(t,y) \di y = \int_\Omega i(t) \di y= \abs{\Omega}\, i(t),
\] 
where $\abs{\Omega}$ denotes the measure of the set $\Omega$.
 
Hence, we now consider the system
\begin{equation}
	\label{eq:ode}
	\begin{cases}
		\dfrac{\di u}{\di t}=pu\Bigl(1-\dfrac{u+i}{K}\Bigr)-\dfrac{\beta}{K} ui-\dfrac{\zeta}{K} uz,\\[8pt]
		\dfrac{\di i}{\di t}=\dfrac{\beta}{K} ui-qi-\dfrac{\zeta}{K} iz,\\[8pt]
		\dfrac{\di z}{\di t}=\alpha i-q_z z +S_0.
	\end{cases}
\end{equation}
The parameter $\alpha$ in Eq. \eqref{eq:ode} corresponds to the parameter $\alpha_z$ of Eqs. \eqref{eq:pde} and \eqref{eq:S} multiplied by the measure of the set $\Omega$ (denoted by $\abs{\Omega}$). 

\begin{figure}
	\centering
	\includegraphics[width=\linewidth]{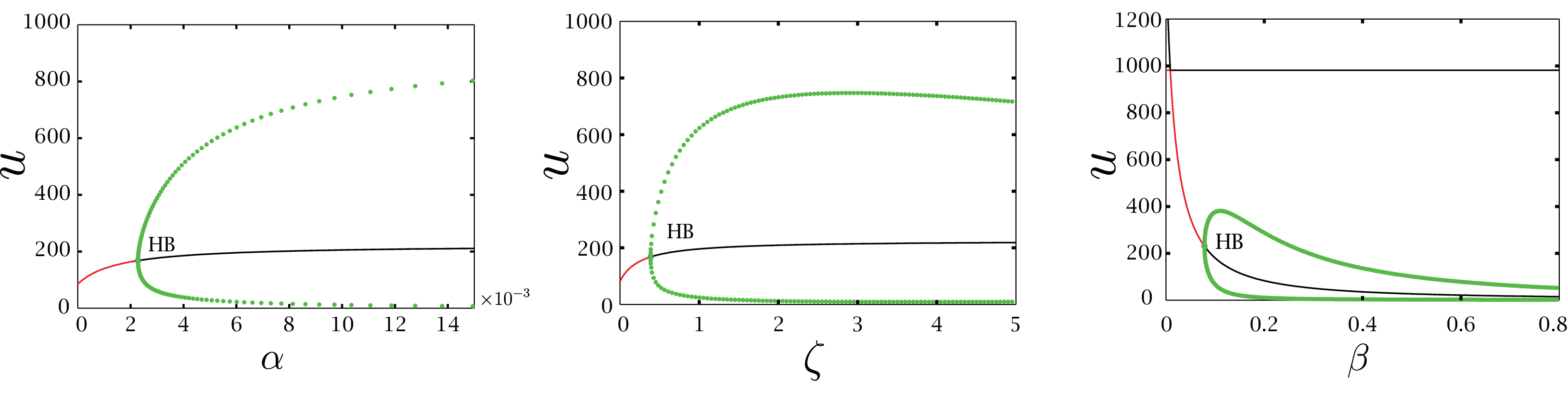}
	\caption{One parameter bifurcations in $\alpha$, $\zeta$ and $\beta$ of Eq. \eqref{eq:ode}, with other parameters as in Table \ref{tab:parameters}. The immune killing rate $\zeta$ has been set to the base value $0.50\;$h$^{-1}$. In order to facilitate comparison with the forthcoming two-dimensional simulations, we set $\alpha=\pi r^2 \alpha_z$ with $r=5\;$mm (corresponding to a late stage of tumour growth). The green dots show the maximum and minimum values of $u$ during the oscillations of the stable limit cycle. The solid lines show the value of the equilibrium of $u$; the line is red if the equilibrium is stable and black if it is unstable. \rev{Hopf bifurcations are denoted by HB.} Observe that for low values of $\beta$ the infection-free equilibrium close to carrying capacity is stable.}
	\label{fig:bif_onepar}
\end{figure}

The equilibria are $(0,0,\frac{S_0}{q_z})$, $(K-\frac{\zeta S_0}{p q_z},0,\frac{S_0}{q_z})$, $(u^*,i^*,z^*)$ and $(0,-\frac{q q_z K}{\alpha \zeta}-\frac{S_0}{\alpha},-\frac{qK}{\zeta})$. The latter exists only for $\alpha,\zeta\neq 0$; it is always negative, so we can neglect it. The third one is defined by the expressions
\begin{equation}
	\label{eq:eq}
	u^*\coloneqq \frac{qK}{\beta}+\frac{\zeta}{\beta} z^*, \qquad
	i^*\coloneqq \frac{K p q_z (\beta-q)-S_0 \beta (\zeta+\frac{p}{\beta}\zeta)}{\beta [q_z(\beta+p)+\alpha (\zeta+\frac{p }{\beta}\zeta)]}, \qquad
	z^*\coloneqq \frac{\alpha}{q_z} i^*+\frac{S_0}{q_z}.
\end{equation}
When $\alpha=S_0=0$, we get $z^*=0$ and for $u$ and $i$ we recover the spatially homogeneous equilibria of Eq. \eqref{eq:l} (model without immune response), namely
\begin{equation}
\label{eq:eq_l}
u^*= \frac{qK}{\beta}, \qquad
i^*= \frac{K p (\beta-q)}{\beta(\beta+p)}.
\end{equation}
As $\alpha$ and $S_0$ increase, $u^*$ increases and $i^*$ decreases. Similarly, when $\zeta=0$, the equilibria are given by Eq. \eqref{eq:eq_l} (although the value of $z$ at the equilibrium may not be $0$).

The \rev{Jacobian matrix computed at the equilibrium point} $(0,0,\frac{S_0}{q_z})$ has eigenvalues $(-q_z, p-\frac{S_0 \zeta}{K q_z}, -q -\frac{S_0 \zeta}{K q_z})$ and the \rev{Jacobian matrix computed at the equilibrium point} $(K-\frac{\zeta S_0}{p q_z},0,\frac{S_0}{q_z})$ has eigenvalues $(-q_z, -p+\frac{S_0 \zeta}{K q_z}, \beta-q -\frac{S_0 (\beta \zeta+p \zeta)}{K p q_z})$. The first equilibrium is stable when
\[
Kpq_z<S_0 \zeta,
\]
corresponding to the situation in which the uninfected cell density of the second equilibrium is negative. This means that the immune system alone may be able to eradicate the tumour without the need for any oncolytic virus. The second equilibrium is stable in the case where neither the first equilibrium is stable nor $i^*>0$. In this case, the oncolytic virotherapy is not effective and the outcome of the therapy depends entirely on the immune response. Let us observe that the density of uninfected cells $K-\frac{\zeta S_0}{p q_z}$ is increasing in the parameters $\zeta$, $S_0$ and decreasing in $p$, $q_z$: while a complete tumour eradication is unattainable, we may still keep the tumour at an acceptable size if the immune response is strong enough.

\begin{figure}
	\centering
	\includegraphics[width=\linewidth]{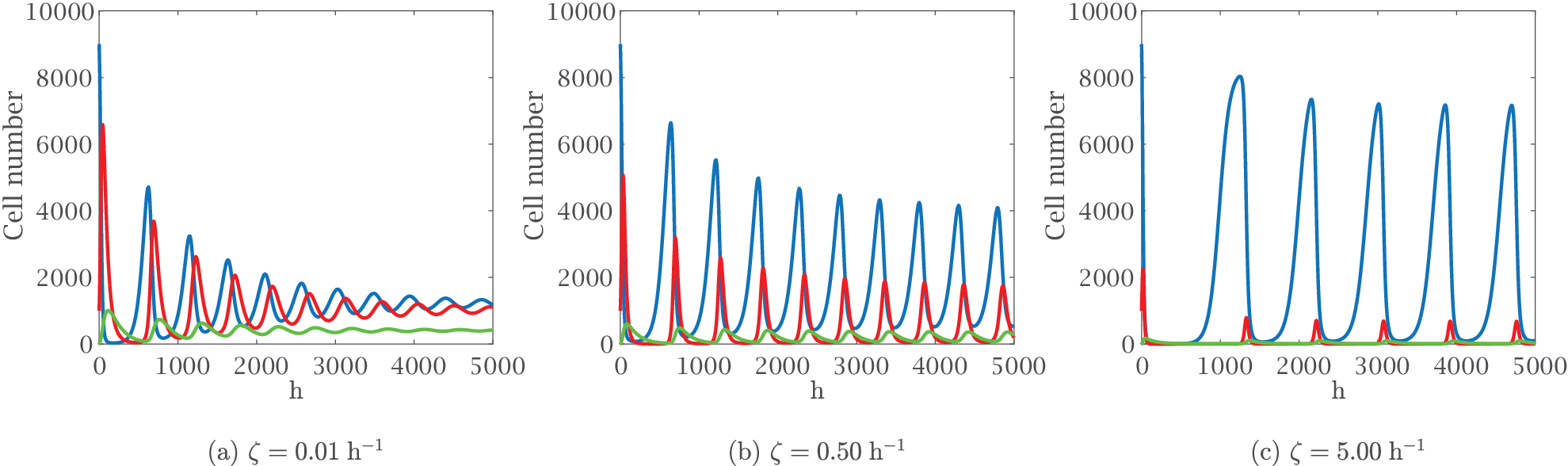}
	\caption[Oscillations of the ODE the interaction of the infection with the immune system]{\rev{Numerical simulation of Eq. \eqref{eq:ode} with the parameters as in Table \ref{tab:parameters} and different values of the immune killing rate $\zeta$. As in Fig. \ref{fig:bif_onepar}, we set $\alpha=\pi r^2 \alpha_z$ with $r=5\;$mm. Uninfected tumour cells are plotted in blue, infected tumour cells in red and immune cells in green. The oscillations become wider as $\zeta$ increases, in accordance with the bifurcation diagram of Fig. \ref{fig:bif_onepar}b.}}
	\label{fig:oscillations}
\end{figure}

The expressions for the eigenvalues of the \rev{Jacobian matrix computed at the third equilibrium point} are more complicated. Numerical simulations show that, in the parameter region where $i^*>0$, either this equilibrium is stable or there appears a stable limit cycle. Fig. \ref{fig:bif_onepar} shows numerical bifurcation diagrams for the parameters $\alpha$, $\beta$ and $\zeta$\rev{. The diagrams were obtained using the software \texttt{auto}, which allows the study of the stability of equilibria and limit cycles through numerical continuation. In all three cases, we observe the appearance of} a Hopf bifurcation; in these continuations, the other parameters of Eq. \eqref{eq:ode} are set to the values of Table~\ref{tab:parameters}.
The size of the oscillations of the limit cycle increases as $\alpha$ and $\zeta$ increase, and decreases as $\beta$ increases. As a consequence, the enhancement of the immune response may significantly decrease the effectiveness of the therapy. However, it is fundamental to also consider that, in some cases, the oscillations have a minimum very close to zero\rev{, as it is clear from the time series of Fig. \ref{fig:oscillations}}: if we take into account a discrete number of cells, they may go extinct when approaching the minimum due to stochastic events and the following regrowth may not take place. Varying the parameter $p$ does not cause a bifurcation of this equilibrium (for parameters within the relevant range), but higher
$p$ values reduce oscillation amplitude during convergence, with some combinations yielding monotone convergence (e.g., for large $p$).

\section{Comparison between agent-based and continuum models}
\label{sec:results}

\begin{table}[t!]
\centering
\scriptsize
\begin{tabular}{lllll}
&\multicolumn{1}{c}{\textbf{Figure}} & \multicolumn{1}{c}{\textbf{Supplementary material}} &  \multicolumn{1}{c}{\textbf{Immune response}} &\multicolumn{1}{c}{\textbf{Infection}}\\
\hline
\\
& Fig. \ref{fig:basics} (solid)        & S1.1     & weak		& no\\
& Fig. \ref{fig:basics} (dashes)          & S1.2     & strong & no\\
& Fig. \ref{fig:basics} (dash-dotted)         & S1.3, S2, S3    & weak & central \\
& Fig. \ref{fig:soloviro_wide}        & S4     & weak		& wide\\
& Fig. \ref{fig:immunoviro_fromstart}         & S5     & strong & central\\
& Fig. \ref{fig:eradication_infection}        & S1.4    & strong & wide \\
& Fig. \ref{fig:eradication_infection}c-d        & S1.5, S7    & strong (delayed) & central \\
& Fig. \ref{fig:violin}         & S6     & strong, high chemotaxis  & wide \\
& Fig. \ref{fig:immunoviro_second}        & S8    & strong & wide, repeated \\
& Fig. \ref{fig:multiple}        & --    & strong & wide, multiple \\
\\
\hline
\end{tabular}
\caption{List of figures and the settings in which they are obtained. Weak and strong immune response refer to the immune killing rate $\zeta$, which takes either of the values listed in Table \ref{tab:parameters}. Central and wide infections are obtained by setting the initial radius of infection $R_i$ either to the reference value or equal to $R_u$.}
\label{tab:fig}
\end{table}

In this section, we compare numerical simulations for the agent-based model and the corresponding system of PDEs. Table \ref{tab:fig} summarises the scenarios explored. The \textsc{Matlab} code is available on GitHub \cite{github26}.

In our simulations we consider a spatial domain $\Omega\coloneqq[-L,L]^2$  with $L=10\;$mm and we adopt zero-flux boundary conditions. We define $\omega$ as in Eq. \eqref{eq:omega} in order to maintain the radial symmetry of the problem, with $R=L$. The initial conditions for Eq. \eqref{eq:pde} are
\begin{equation}
	\label{eq:initial}
	u_0(\vec{x})=
	\begin{cases}
		0.9\ K \quad &\text{for } \abs{\vec{x}}\leq R_u, \\
		0\ \quad &\text{for } \abs{\vec{x}}> R_u, 
	\end{cases}
	\qquad
	i_0(\vec{x})=
	\begin{cases}
		0.1\ K \quad &\text{for } \abs{\vec{x}}\leq R_i, \\
		0 \quad &\text{for } \abs{\vec{x}}> R_i,
	\end{cases}
\end{equation}
where $R_u$ and $R_i$ are respectively the initial radius of uninfected and infected cells; initial conditions for $z$ and $\phi$ are 0 across the whole domain. Analogous initial conditions are used for the agent-based model, with the only difference that cell numbers rather than cell densities are considered. The reference case with $R_u>R_i$ corresponds to the intratumoral injection of the virus \cite{jin21}. On the other hand, in some simulations we assume that $R_u=R_i$, which corresponds to an infection of the whole domain: since we consider a tumour that can be infiltrated by the immune system without major obstacles, it is reasonable to assume that this could be obtained with an intravenous administration of the virus \cite{jin21}. 

These assumptions guarantee radial symmetry in the continuum formulation, thus justifying the use of Eq. \eqref{eq:pderad}. In contrast, the two-dimensional agent-based simulations do not display perfect radial symmetry due to stochastic fluctuations. Nevertheless, the comparison remains meaningful: the formal derivation in Appendix \ref{app:derivation} shows that, on average, the evolution of the agent-based model---though not strictly symmetric---corresponds to that of the radially symmetric continuum limit.

We use the parameters listed in Table \ref{tab:parameters}, whose choice is motivated in Appendix \ref{app:num}. We assume $\zeta=0.50\,$h$^{-1}$ as the base immune killing rate of tumour cells; an enhancement of the immune system (e.g., due to immune checkpoint inhibitor therapy) is then modelled by increasing it to $\zeta=5.00\,$h$^{-1}$.

The discussion of Section \ref{sec:ode} suggests that some parameter ranges may lead to persistent oscillations in the centre of the domain.
Oscillations may bring the cell density to such low levels that the agents go extinct. When this happens, it is convenient to perform a high number $M$ of simulations and quantify the \textit{tumour control probability} at time $t_n=\tau n$ as
\begin{equation}
\label{eq:tcp_dis}
\text{TCP}(\tau n)\coloneqq \frac{1}{M} \abs*{\Set{m |\sum_h U_h^n=0 \text{ in simulation } m}},
\end{equation}
where $\tau$ is the time discretisation for the agent-based model and $\abs{\cdot}$ denotes the cardinality of the set. In other words, TCP$(\tau n)$ is the ratio of simulations that result in the extinction of the uninfected tumour cells in the interval $[0,\tau n]$. As infected cancer cells cannot proliferate, their survival does not affect the treatment outcome. Note that the dependence on time of the right-hand side of the equation is implicit in $U_h^n$, which denotes the cell number at a given position and time $t_n=\tau n$. This definition is coherent with previous papers on the subject (see Ref. \cite{gong13} and the references therein).

The continuum model, on the other hand, cannot model cancer eradication, as the zero-solution is unstable and, as a consequence, it can only be approximated but never attained. Nevertheless, it makes sense to assume that the number of surviving uninfected cells follows a Poisson distribution and define 
\begin{equation}
\label{eq:tcp_con}
\text{TCP}(t)\coloneqq \exp\left[- 2\pi\int_0^{R} u(t,r)r\di r\right].
\end{equation}
In Ref. \cite{gong13}, the Poissonian TCP defined as above was compared with more complex expressions and it was found to perform equally well in the case of radiotherapy. It is therefore interesting to perform a similar comparison in the context of our model.

Infected cells are much more likely to be eradicated than uninfected cells since even a small population of the latter tends to regrow. In several cases it is therefore useful to consider an analogous probability to quantify the extinction of infected cells, which we denote \textit{infection control probability}. It is defined as 
\begin{equation}
\label{eq:icp_dis}
\text{ICP}(\tau n)\coloneqq \frac{1}{M} \abs*{\Set{m |\sum_h I_h^n=0 \text{ in simulation } m}},
\end{equation}
in the case of the agent-based model and as
\begin{equation}
\label{eq:icp_con}
\text{ICP}(t)\coloneqq \exp\left[- 2\pi\int_0^{R} i(t,r)r\di r\right].
\end{equation}
in the case of the continuous model.

\subsection{Partially effective treatments}

\begin{figure}
	\centering
	\includegraphics[width=0.9\linewidth]{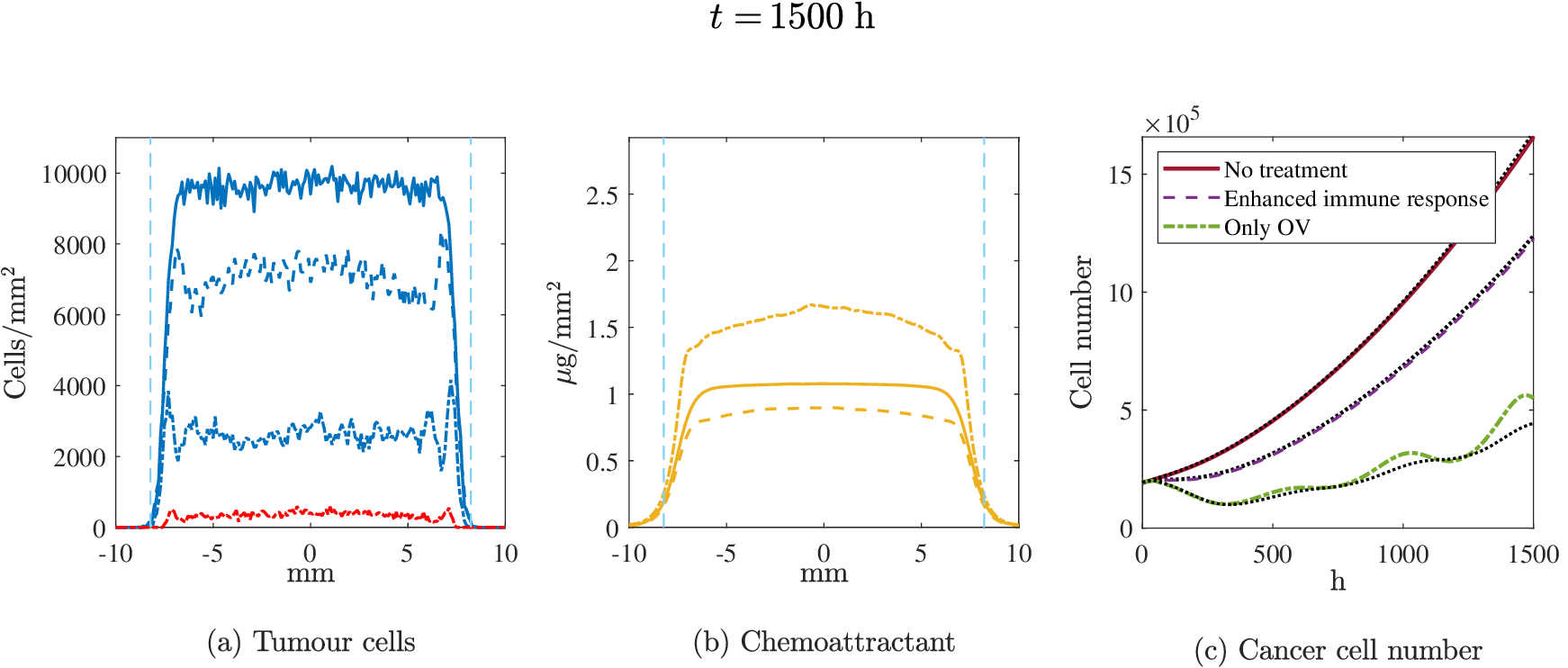}
	\caption{Comparisons of the scenarios with weak immune response and no infection (solid lines, $\zeta=0.50\,$h$^{-1}$ and $R_i=0$), strong immune response and no infection (dashed lines, $\zeta=5.00\,$h$^{-1}$ and $R_i=0$), weak immune response and central infection (dash-dotted lines, $\zeta=0.50\,$h$^{-1}$). All the other parameters take the values given in Table \ref{tab:parameters}. The results of the agent-based model are averaged over ten simulations. Panels (a-b) represent the horizontal section of the domain $[-L,L]\times\{0\}$: uninfected cells are in blue, infected cells in red (only shown with dash-dotted lines, as in the other cases there is no infection) and chemokines in yellow. The vertical blue dashed lines represent the expected positions of the uninfected invading front, travelling at speed $2\sqrt{D_u p}$. Panel (c) shows the sum of tumour cells over time. The results of the agent-based model in the three scenarios are represented with solid, dashed and dash-dotted lines (as explained in the legend). The results of the continuum models are represented with dotted lines in all the three cases, as there is an excellent agreement with the stochastic counterparts.}
	\label{fig:basics}
\end{figure}

We first describe some simple settings, in order to better understand the basic interactions between the tumour, the infection and the immune system. Fig. \ref{fig:basics} represents the following scenarios:
\begin{enumerate}
\item solid lines: weak immune response, no infection;
\item dashed lines: strong immune response, no infection;
\item dash-dotted lines: weak immune response, central infection by oncolytic virus.
\end{enumerate}
In all these cases, there is an excellent qualitative agreement between numerical solutions of the system of PDEs \eqref{eq:pderad} and a single simulation of the agent-based model. When the infection is present, the agreement is also quantitative. On the other hand, the number of immune cells involved in the cases without infection is so low that stochastic fluctuations cannot be neglected; hence, the quantitative difference between the two modelling approaches is significant; however, it is enough to consider the average over ten simulations to obtain an improved quantitative agreement. 

In the first case, the tumour grows almost up to carrying capacity and the immune response is very ineffective. The maximum density of the chemoattractant stabilises around a value slightly larger than $1\,\mu$g/mm$^2$, which is slightly more than a third of $\phi^*$: this is enough to guide immune cells inside the tumour and Fig. S1.1 in the electronic supplementary material clearly shows that the internal density is above the base value $S_0/q_z$. The presence of immune cells decreases the tumour cell density to approximately $K-\frac{\zeta}{p} z(t,\vec{x})$, which in this case is only slightly below $K$. When the immune killing rate $\zeta$ is increased, clearly, the tumour cell density decreases. Nevertheless, this does not significantly increase the effectiveness of the immune response: indeed, fewer tumour cells secrete less chemoattractant and the chemotactic component of the immune cell movement is weaker than before, leading to a lower density of immune cells inside the tumour (see Fig. S1.2 in the electronic supplementary material). This is in line with the empirical observation that immunotherapy alone usually cannot eradicate cold tumours \cite{galon19}. Indeed, an increased an increased arrival rate of immune cells to the tumour microenvironment (i.e., a higher value of the parameters $\alpha_z$ and $S_0$) or a more effective immune response may not be able to eradicate the mass completely, as a few tumour cells do not secrete enough chemoattractant to guide immune cells.

These results motivate the use of oncolytic virotherapy to reduce the tumour load. The dynamics in this reference situation are very similar to the ones described in Appendix \ref{subsec:bmb}. All the formulas related to the invasion speeds are not affected by the immune system, since the linearised equations do not change. On the other hand, the central densities are affected by immune cells: the exact values are hard to predict due to the presence of the chemotactic term, but we can verify that uninfected cell density is higher than in the absence of immune response and infected cell density is lower, as predicted by the analysis of Eq. \eqref{eq:ode}. The chemokines' density in the central region is $\phi^*$ in the first phases of the infection and  this guarantees a high immune infiltration of the tumour; it later decreases to approximately $1.\rev{5}\;\mu$g/mm$^2$ due to the reduced amount of cells, which is still higher than the values we observed in the absence of viral infection and enough to guide a high number of immune cells towards the tumour. The highest peak of immune cell density corresponds at all times to the invading front of infected cells, as this is the region of the steepest gradient of chemokines. Overall, the total cancer cell number is significantly lower than what we observed without viral infection, as shown in Fig. \ref{fig:basics}c, meaning that the treatment is at least partially effective.

Overall, the immune response reduces the efficacy of the infection: in comparison to the results sketched in \ref{subsec:bmb}, the invasion speed of the infected cells is slightly lower because of the lower uninfected cell density; furthermore, the equilibrium value of uninfected cells is higher and the one of infected cells is lower (see Eq. \eqref{eq:eq} and the comments thereafter). Despite this difference, most of the conclusions of Ref. \cite{morselli23} to optimise virotherapy  (summarised in \ref{subsec:bmb}) remain valid: as it is clear from Eq.~\eqref{eq:eq}, the efficacy of the infection increases as $\beta$ increases and $q$ decreases.

\subsection{Emergence of oscillations}

\begin{figure}
	\centering
	\includegraphics[width=\linewidth]{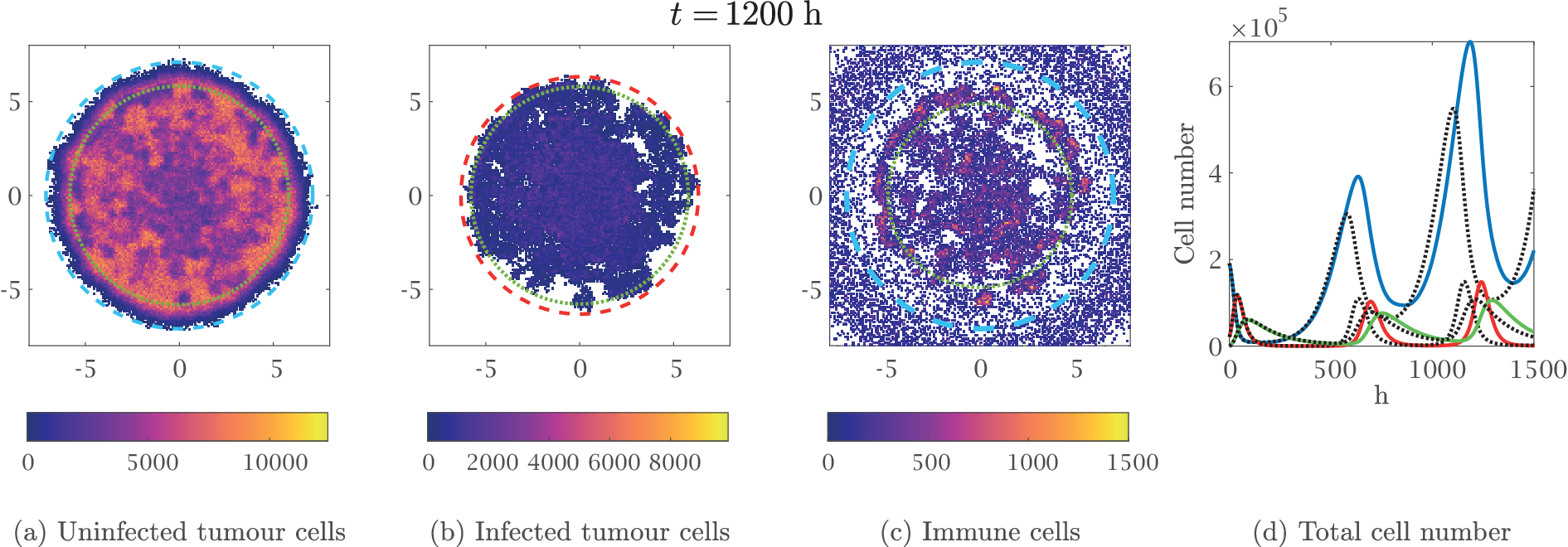}
	\caption{A \rev{single} numerical simulation of the agent-based model with the parameters given in Table \ref{tab:parameters}, $\zeta=0.50\,$h$^{-1}$ and wide oncolytic viral infection (i.e., $R_i=R_u$). The dashed cyan circles in panels (a) and (c) represent the expected positions of the tumour invading front, travelling at speed $2\sqrt{D_u p}$. The dotted green circles in panels (a), (b) and (c) represent the internal minimum of the numerical solution of Eq. \eqref{eq:pderad}. The dashed red circle in panel (b) represents the front given by the numerical solution of Eq. \eqref{eq:pderad}. In panel (d), solid lines refer to the agent-based model (uninfected, infected and immune cells are represented respectively in blue, red and green) and dotted lines refer to the continuum model. In all the cases the maximum of the axes and the colorbars correspond to the maximum over time of the quantity plotted.}
	\label{fig:soloviro_wide}
\end{figure}

The discussion of Section \ref{sec:ode} suggests that some parameter ranges may lead to persistent oscillations in the centre of the domain. We should also take into account that these oscillations may be biologically relevant even if they converge towards a stable equilibrium, as the convergence may be very slow. Fig. \ref{fig:soloviro_wide} along with the video accompanying it (see electronic supplementary material S4), indeed shows an example of this situation: the only difference with respect to the reference case is the initial condition $R_i=R_u$ (i.e., the initial viral injection covers the whole domain); we, therefore, expect the same asymptotic behaviour of the reference situation, but up to time $t=1500\;$h the difference between the scenarios of Figs. \ref{fig:basics}c (dash-dotted lines) and \ref{fig:soloviro_wide}d is very significant. This can be explained by noting that the initial number of infected cells is higher; hence, more immune cells are involved, which causes wider oscillations. It is interesting to observe that the oscillations of the agent-based model are delayed with respect to the ones of the continuum model (this is easy to see from the total number of cells): when the infected cell density is very low, stochasticity becomes relevant and, in some regions, infected cells go extinct; hence, the following infected cell regrowth is at first inhomogeneous and it takes some time to diffuse in the whole domain. Despite the difference of the spatial pattern, the area occupied by the tumour is the same in both cases (as shown by the red circle in Fig. \ref{fig:soloviro_wide}a, representing the theoretical position of the invading front) and the infection manages to reach the boundary of the tumour in a short time (as shown by the cyan circle in Fig. \ref{fig:soloviro_wide}b, representing the front of infected cells for the PDE). The spatial inhomogeneities quickly disappear, and the delay in the oscillations is the main difference between the two modelling approaches. We remark that these spatial inhomogeneities are amplified by the dynamics and similar patterns would also arise in the two-dimensional non-radial continuum model under stochastic perturbations. 

\begin{figure}
	\centering
	\includegraphics[width=0.8\linewidth]{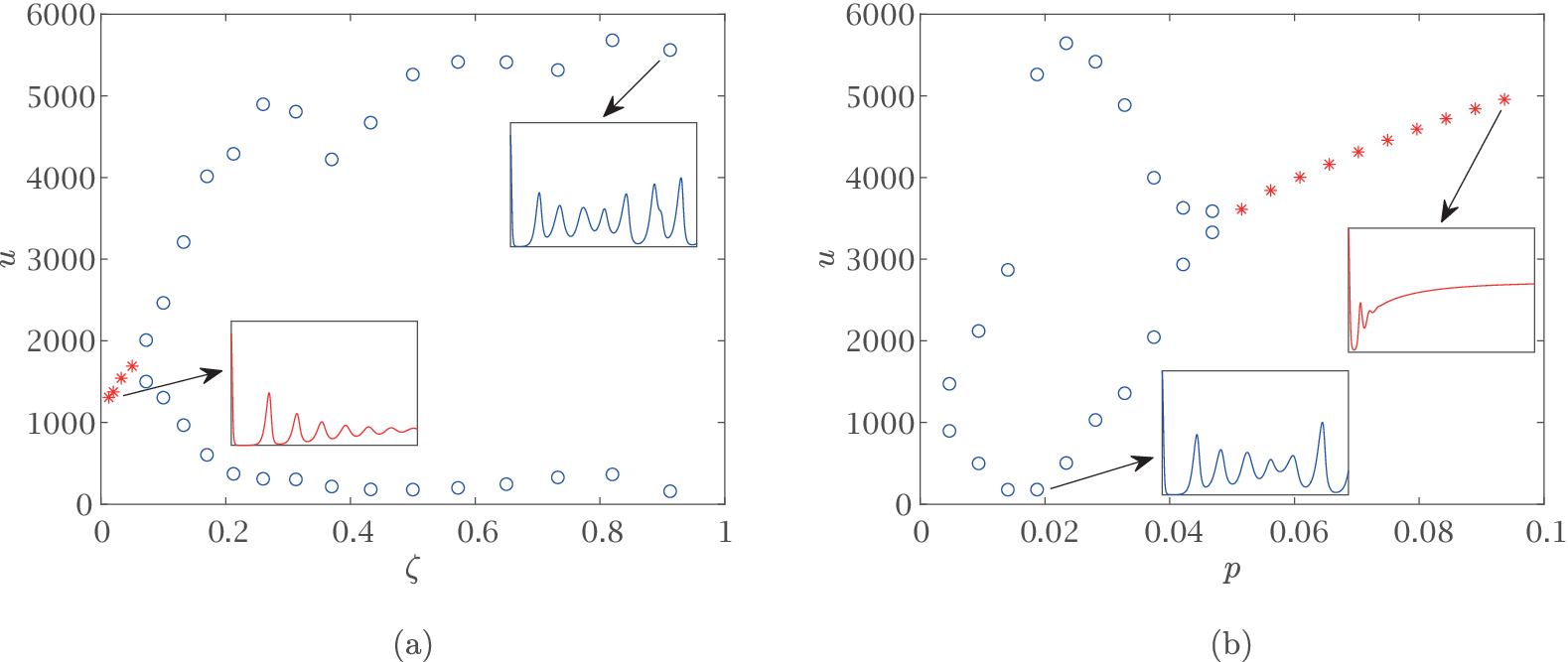}
	\caption{Oscillations at the origin from numerical simulations of the PDE for different values of $\zeta$ and $p$, with the other parameters as in Table \ref{tab:parameters}, $\zeta=0.50\;$h$^{-1}$ in panel \rev{(b)} and the chemotactic coefficient $\chi$ reduced to $1.65\times 10^{-2}\;$mm$^2$/h (i.e., one-tenth of the reference value) in order to guarantee numerical stability for all the parameter values under investigation. The simulations are run respectively until $4000\;$h and $3500\;$h on circular domains of radius $20\;$mm and $25\;$mm in order to avoid boundary effects. The blue dots show the maximum and minimum values of $u(\cdot,0)$ during the oscillations after $t=3000\;$h (when present). The red stars show the value of $u$ in the centre at the last time in cases where oscillations dampen significantly.}
	\label{fig:bif_p}
\end{figure}

Different parameter ranges may result in wider oscillations, which cause extinction in the agent-based model.
It makes sense to take the bifurcation diagrams of Section \ref{sec:ode} as a starting point and study whether system \eqref{eq:pderad} has the same behaviour as parameters vary. Fig.~\ref{fig:bif_p} indeed shows that, as $\zeta$ increases, the oscillations that appear show an increasing maximum value and a decreasing minimum value, suggesting that, in the agent-based model, both uninfected and infected cells should become close to eradication. The influence of the parameter $p$ is more interesting: the stability of the equilibrium $(u^*,i^*,z^*)$ appears independent of its value, but the size of the oscillations during the convergence decreases as $p$ increases (see Fig. \ref{fig:bif_p}); this leads to the counter-intuitive result that a fast-growing tumour has a more predictable behaviour under therapy than one with slower growth.

\subsection{Eradication of infection}

Oscillations may bring the cell density to such low levels that the agents go extinct, even though the continuum model predicts recurrence. Infected cells are much more likely to be eradicated than uninfected cells since even a small population of the latter tends to regrow; this means that, in practice, most of the time after the first oscillation, the infection is eradicated and the tumour keeps growing as it would do in the absence of virotherapy. 
This motivates our interest in exploring different parameter ranges. While the increase of the infection rate $\beta$ may be challenging to implement biologically, the enhancement of the immune response appears more feasible, for example, through the use of immune checkpoint inhibitors \cite{iwai02} or other immune boosting techniques (T-cell transfer, immune system modulators, etc.), as in Refs. \cite{he23,krabbe21,wing18}. 

\begin{figure}
	\centering
	\includegraphics[width=\linewidth]{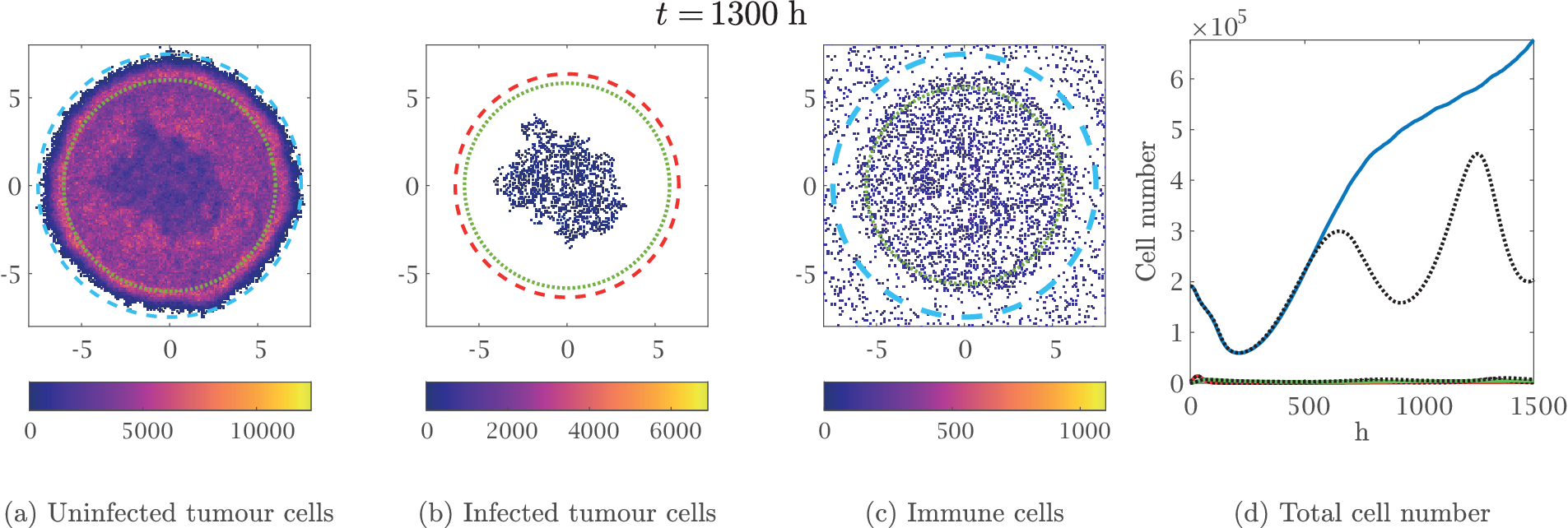}
	\caption{A \rev{single} numerical simulation of the agent-based model with the parameters given in Table~\ref{tab:parameters}, $\zeta=5.00\,$h$^{-1}$ and central oncolytic viral infection. All the graphical elements have the same meaning as Fig. \ref{fig:soloviro_wide}.
}
	\label{fig:immunoviro_fromstart}
\end{figure}

Increasing the immune killing rate of cancer cells to $5.00\;$h$^{-1}$ gives the simulations in Fig. \ref{fig:immunoviro_fromstart}, along with the video accompanying it (see electronic supplementary material  S5). The agent-based model and the continuum model show an excellent quantitative agreement up to approximately $t=490\;$h, when we observe the recurrence of the infection only in the continuum model (as shown by the internal circles); this also stimulates again the immune system. In the agent-based model, infected cells are extinct in most of the domain and the subsequent infection remains confined in the centre of the tumour. Nevertheless, the ICP computed over 100 simulations is 0: the immune response is strong enough to keep the infection under control, but too weak to eradicate it. 

\begin{figure}
	\centering
	\includegraphics[width=\linewidth]{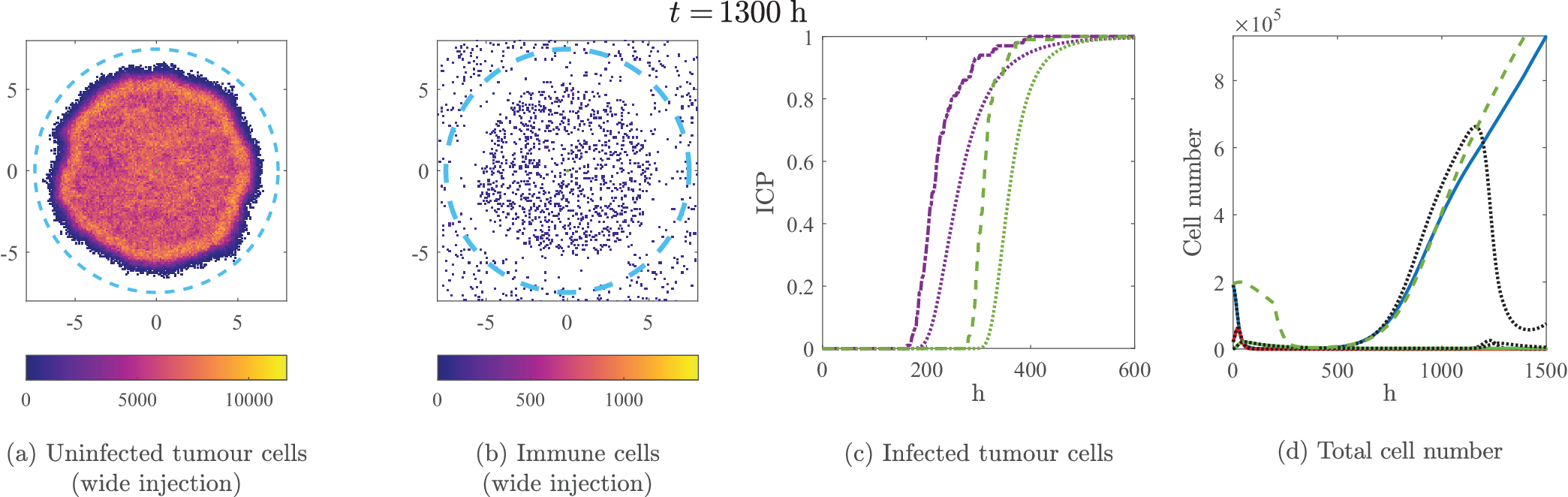}
	\caption{(a)-(b) A \rev{single} numerical simulation of the agent-based model with the parameters given in Table \ref{tab:parameters}, $\zeta=5.00\,$h$^{-1}$ and wide oncolytic viral infection (i.e., $R_i=R_u$); the radius of infection is the only difference with respect to Fig. \ref{fig:immunoviro_fromstart}. All the graphical elements have the same meaning as Fig. \ref{fig:soloviro_wide}. (c)-(d) Comparison between an initially wide infection with an enhanced immune response from the beginning (purple line in panel (c) and most lines in panel (d)) and central initial infection and a delayed enhancement of the immune response (green lines). Solid and dashed lines refer to the agent-based model; dotted lines refer to the continuum model. }
	\label{fig:eradication_infection}
\end{figure}

A stronger immune response can be obtained by considering an infection that is spread in the whole tumour, so that the higher number of infected cells increases the influx of immune cells. We first achieved this by considering an initially wide infection with an enhanced immune response from the beginning, as shown in Fig. \ref{fig:eradication_infection}a-b-d and purple lines in Fig. \ref{fig:eradication_infection}c, as well as electronic supplementary material S1.4. We then considered a central initial infection and a delayed enhancement of the immune response, as shown in Fig. \ref{fig:eradication_infection}c-d, as well as electronic supplementary material S1.5 and S7. In the second case, the immune killing rate is increased at time $t=200\;$h, corresponding to the moment in which the infected front reaches the tumour boundary. The dynamics are very similar in the two cases, with a significant initial tumour reduction due to the combination of the infection with the strong immune response. We performed again $100$ simulations for each scenario and observed in all of them the eradication of the infection. It is interesting to observe a good qualitative agreement of the two ICPs computed according to Eqs. \eqref{eq:icp_dis} and \eqref{eq:icp_con}: the continuous model is indeed able to predict a faster eradication in the first of the two cases. Nevertheless, we should also remark that there is a consistent underestimation of the continuous ICP with respect to the one resulting from the agent-based model. As time passes, the number of immune cells in the domain decreases; the few remaining uninfected cells left secrete too little chemoattractant to guide the immune action; hence, they start to regrow. The TCP is $0$ for both models, meaning that stochasticity does not appear to play a key role in the process, despite the low cell numbers involved. Overall, the regrowth is only slightly delayed with respect to the continuum model (see also electronic supplementary material S1.5).

\begin{figure}
	\centering
	\includegraphics[width=\linewidth]{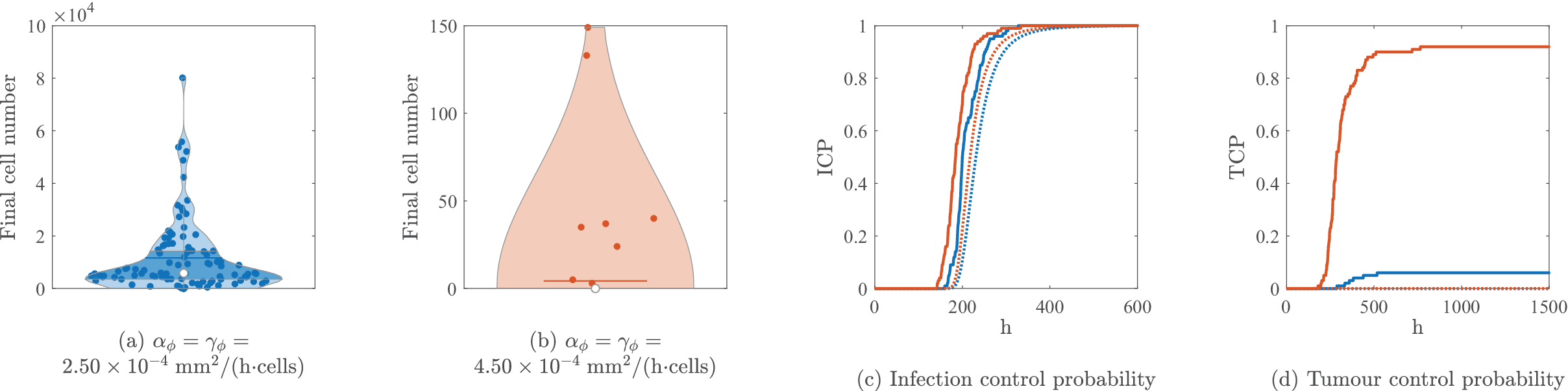}
	\caption{Tumour eradications of the agent-based model with the parameters given in Table \ref{tab:parameters}, $\zeta=5.00\,$h$^{-1}$, wide oncolytic viral infection (i.e., $R_i=R_u$) and different chemoattractant secretion rates, represented in blue and red. (a)-(b) Violin plots of the final cell number at time $t=1500\;$ obtained from one hundred simulations. The dots show the single results, not shown when they are at $0$. The white dots show the median and the blue horizontal lines show the average. The dark blue areas in panel (a) show the region between the first and the third quartile; this is not shown in panel (b), as both quartiles coincide with zero. (c)-(d) ICP and TCP for the previous situations (blue lines in the case of panel (a) and red lines in the case of panel (b)). Solid and dashed lines refer to the agent-based model; dotted lines refer to the continuum model.}
	\label{fig:violin}
\end{figure}

It is interesting to observe that the increase of immune cell density at later times (due, for example, to CAR-T therapy \cite{he23}) is not enough alone to eradicate the few tumour cells left, as T-cells are unable to effectively detect tumour cells (not shown). On the other hand, for higher chemoattractant secretion rates, the immune system may be able to keep the few surviving cancer cells under control and, in some cases, completely eradicate the tumour. In this setting, we observe a significant effect of stochasticity due to the very low uninfected cell number involved. 
When $\alpha_\phi$ has the reference value and $\gamma_\phi=\alpha_\phi$ (Fig. \ref{fig:violin}a and blue lines in Fig. \ref{fig:violin}c-d), tumour eradication is observed only in a few simulations and, in some others, the tumour appears under control for a long time (see supplementary material S6 for an example of this); nevertheless, the majority of the simulations show recurrence. When the secretion rates are even higher (Fig. \ref{fig:violin}b and red lines in Fig. \ref{fig:violin}c-d), the immune system appears able to eradicate the tumour in the wide majority of cases. The eradication of the infection is captured extremely well by the Poissonian ICP, as shown in Fig. \ref{fig:violin}c. On the other hand, the Poissonian TCP is very close to $0$ in both cases and appears totally unable to predict the tumour extinction events of the agent-based models. 
A possible explanation for this discrepancy is that localised cell clusters are more effectively cleared by the immune system than wider, spatially homogeneous populations with a lower density. In the agent-based models, stochastic fluctuations during the infection naturally generate such small clusters, whereas the continuum model represents only averages and therefore cannot reproduce this effect. As a result, the total cell number is comparable in both models during the decrease, but the actual spatial configuration significantly affects the subsequent dynamics. Overall, randomness plays a very important role in therapeutic outcomes. We expect that introducing stochasticity into the continuum model would reconcile the two approaches, although determining an appropriate formulation is nontrivial.


These scenarios could be interpreted as the situation of a tumour with a very high mutational burden that is well recognised by the immune cells in the area, despite not stimulating the immune system by itself (see, for example, the review Ref. \cite{odonnel18} and type 1 tumours described there); it is therefore highly likely that such a tumour never reaches a significant size, as any attempt to grow would be immediately stopped by the immune cells already present in the area. Our results show that immunovirotherapy could be efficient in the few cases in which this kind of tumour evades immune control. However, a direct increase in the chemokine secretion of uninfected tumour cells poses several challenges in the clinical settings \cite{abdulrahman24}. We remark that, in all these situations, the agent-based model differs significantly from the continuous model, which always shows tumour relapse.

\subsection{Different treatment protocols}

\begin{figure}
	\centering
	\includegraphics[width=0.8\linewidth]{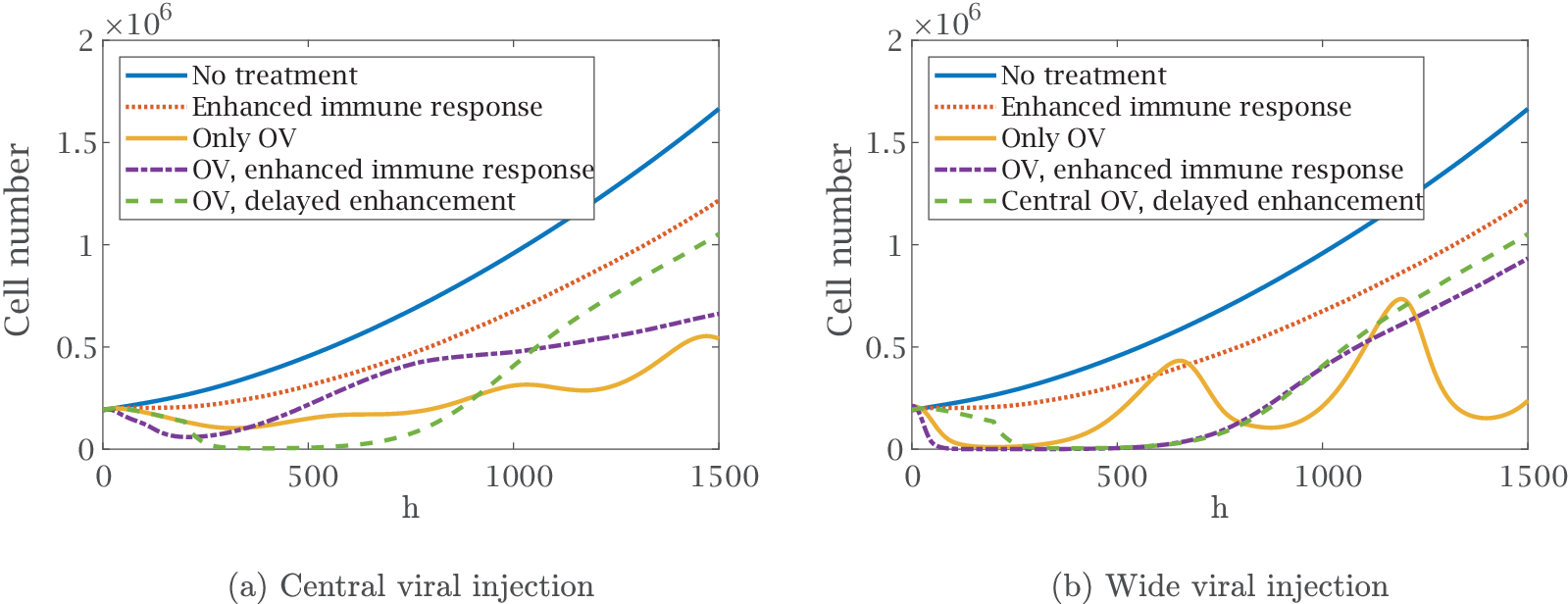}
	\caption{Comparison between the sum of tumour cells obtained as the average of five agent-based simulations in different regimes in case of central viral injection (a) and wide viral injection (b). It is here evident that the enhancement of the immune system during virotherapy may worsen the outcome. The delay of the enhancement in the case of central injection is very similar to the case of a wide injection with the immune system enhanced from the beginning.}
	\label{fig:comparison}
\end{figure}

\begin{figure}
	\centering
	\includegraphics[width=\linewidth]{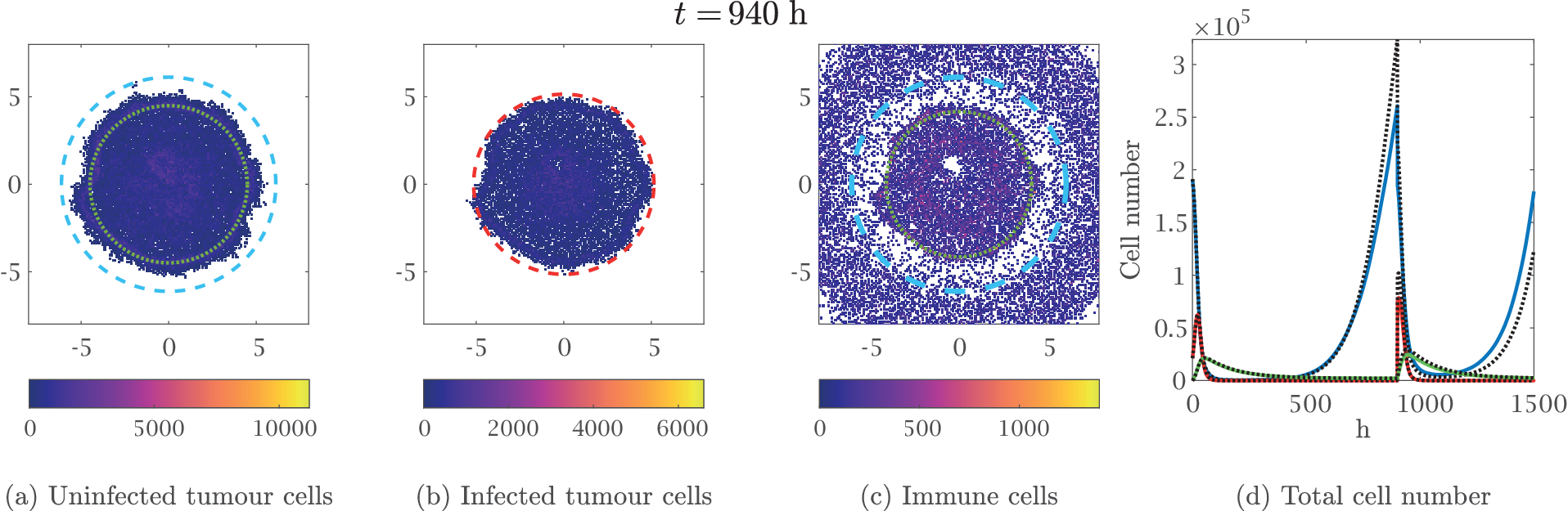}
	\caption{A \rev{single} numerical simulation of the agent-based model with the parameters given in Table \ref{tab:parameters}, $\zeta=5.00\,$h$^{-1}$ and wide oncolytic viral infection (i.e., $R_i=R_u$); a second viral injection is performed at time $T_1=900\;$h, infecting $30$\% of cells everywhere. All the graphical elements have the same meaning as Fig. \ref{fig:soloviro_wide}. 
	}
	\label{fig:immunoviro_second}
\end{figure}

Fig. \ref{fig:comparison} summarises the total number of cells of the agent-based model in the different scenarios analysed so far, clearly showing that the combination of treatments may worsen the outcome.
The best result that we have achieved so far without changing the chemoattractant secretion rate is a temporary tumour remission in the situation of an initial infection that affects the whole area of the tumour when the immune response is enhanced (green lines of Fig. \ref{fig:comparison}a-b and purple line of Fig. \ref{fig:comparison}b). 

Our goal is now to exploit this temporary remission and try to achieve a better therapeutic outcome, keeping in mind that a persistent infection reduces the tumour burden for indefinitely long times (as in the yellow lines of Fig. \ref{fig:comparison}). A possible solution would be to have a second viral injection when the tumour starts to relapse: this requires to define the timing and location for the injection. As the spatial configuration of the tumour during the remission is very sparse, it is hard to decide a priori the location for a localised infection. We overcome the issue by assuming for simplicity that the virus may easily reach any area of the tumour and so we consider a wide infection. Our focus is then the timing for the second viral injection: on the one hand, we do not want to wait until the tumour is too big as this is inconvenient for the patient; on the other hand, if we do it early when the number of cells is too low, the infection quickly dies away.

Fig. \ref{fig:immunoviro_second}, along with the video accompanying it (see electronic supplementary material  S8), shows the scenario in which the immune system is enhanced all the time and the first wide injection at time $T_0=0\;$h is followed by a second one at time $T_1=900\;$. We assume that this second viral injection causes $30\%$ of the tumour cells to become infected, irrespective of their location. The tumour is now kept under control for a longer period of time with respect to the cases shown in Fig. \ref{fig:eradication_infection}c-d. However, we always observe a later recurrence. 


This suggests that a good therapeutical approach could be to perform periodic repeated viral injections: let us now focus on optimising the schedule. Fig. \ref{fig:immunoviro_second} shows a very good quantitative agreement between numerical solutions of the system of PDEs \eqref{eq:pderad} and single numerical simulations of the agent-based model. We exploit this fact to reduce our attention to the continuum model and simulate an automatic viral injection when the uninfected cell count reaches a fixed threshold $U_{\text{threshold}}$ decided a priori. We assume that the first viral injection happens at time $T_0=-500\;$h. As our goal is the optimisation of treatment at long times, we do not need to focus on the initial transient dynamics up to time $t=0\;$h, which are common for all our treatment strategies. The following injections happen at times 
\begin{equation*}
T_j\coloneqq \inf \Set{t>T_{j-1} |  2\pi\int_0^{R} u(t,r)r\di r \geq U_{\text{threshold}}}.
\end{equation*}
We again assume that the additional viral injection causes $30\%$ of the tumour cells to become infected, irrespective of their location. The second injection time has the additional constraint $T_1>0$; this does not affect the outcome for the thresholds that we consider.

\begin{figure}
	\centering
	\includegraphics[width=0.8\linewidth]{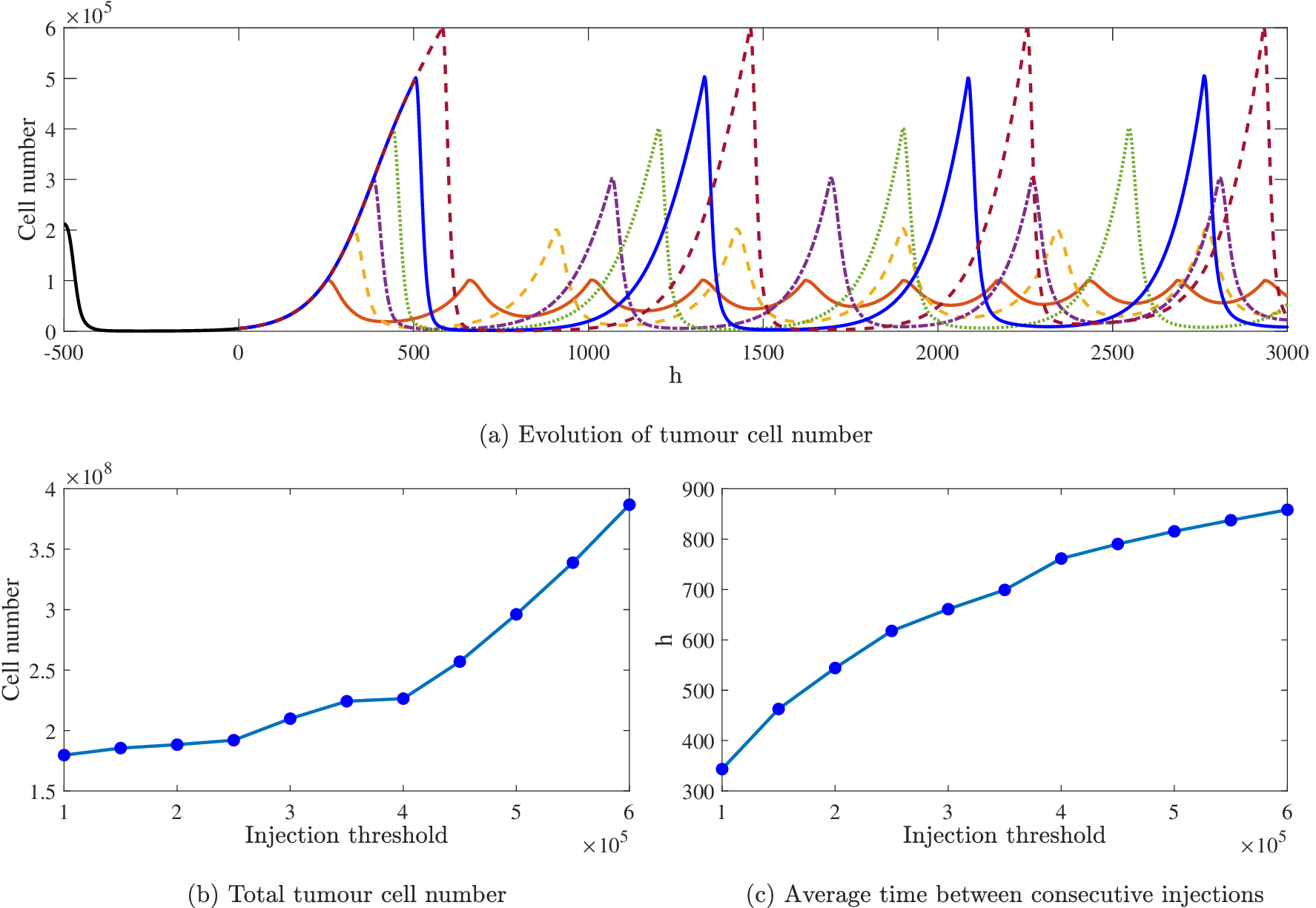}
	\caption{Numerical solutions of Eq. \eqref{eq:pderad} with the parameters given in Table \ref{tab:parameters}, $\zeta=5.00\,$h$^{-1}$ and wide oncolytic viral infection (i.e., $R_i=R_u$); after the first $500\;$h, a new wide viral injection is performed in the whole tumour as soon as the cell count reaches a given threshold (decided a priori) and as a result $30$\% of cells in every location become infected. Panel (a) shows the total tumour cell numbers for some thresholds. Panels (b) and (c) show respectively the total tumour cell number from $t=0\;$h to $t=3000\;$h and the average time between two consecutive injections for different values of the threshold at which the injection is performed.}
	\label{fig:multiple}
\end{figure}

Fig. \ref{fig:multiple} shows that, as $U_{\text{threshold}}$ increases, the minimum tumour size achieved decreases, but the total area under the curve increases. An ideal treatment would require very frequent viral injections, but its implementation in real life may be inconvenient. Nonetheless, for some chronic conditions where the patient requires lifelong, periodic monitoring, this approach should not be completely discarded.

\section{Conclusions}
\label{sec:conclusions}

A minimal, hybrid discrete-continuum model for the interactions between tumour cells, oncolytic viruses and the immune system has been developed. The deterministic continuum counterpart is formally derived and the numerical results of the two approaches are compared. The main assumption is that the tumour under investigation is immunologically cold (i.e., its immunogenicity is very low) and the viral infection stimulates an immune response.

The continuum model is an excellent approximation of the underlying microscopic model in several cases. This allows us to improve our understanding of the therapeutic outcome in different settings, relying on some analytical insights coming from the analysis of the nonspatial model and performing extensive numerical simulations in a reasonable amount of time. On the other hand, in some situations, we observe significantly different behaviours between the two models. The main explanation for this is the appearance of oscillations that bring the cell density to very low levels: the continuum model may then exhibit a quick regrowth, whereas in the same conditions the agent-based model can exhibit extinctions for low population counts. This extinction is more likely to happen for infected cells, since uninfected cells have the ability to regrow; therefore, it appears as a major obstacle to the effectiveness of immunovirotherapy. The bifurcations of the corresponding ODE mirrors the oscillations of the spatial continuum model, which are often associated to discrepancies between the agent-based and the continuum model. We remark on the importance of taking into account also the transient behaviour of the system, since oscillations may dampen on a time scale much longer than the biologically meaningful one.
Even though stochasticity plays a key role in the process, the infection control probability (ICP) for the continuum model characterises well the extinction of infected cells. On the other hand, the tumour control probability (TCP) fails to capture the eradication of the tumour. 

Our results show that, according to the continuum model, any immune response has the tendency to decrease the effectiveness of the virotherapy. This holds true for the agent-based model whenever oscillations are absent or too weak to drive the uninfected cell number very low; in the latter situation, the partial extinction of the solely infected populations may result in a complete failure of virotherapy. On the other hand, stronger oscillations are sometimes able to lead all the cancer cells close to extinction in the individual-based model. This happens when the infection has the possibility to propagate in the whole tumour before the enhancement of the immune system; hence, it can only be achieved if the time and location of the therapies are correctly calibrated. Fig. \ref{fig:comparison} clearly shows that the combination of treatments may worsen the outcome; this result is in line with several experimental evidence \cite{filley17}. \rev{At the final time of the simulations $t=1500\;$h, the best outcome is achieved through virotherapy alone and any kind of immune enhancement worsens it. On the other hand, other treatment protocols show much more promising results at earlier times.} Even though our model shows recurrence of the tumour at later times in most cases, such a low number of cells suggests that it would be possible to completely eradicate the tumour (e.g., by implementing an additional therapy) or at least to keep it under control (e.g., by repeated viral injections, as shown in Fig. \ref{fig:multiple}).

There are several factors that may hinder the success of immunovirotherapy predicted by our model. First, we assume that immune cells kill uninfected and infected cells at exactly the same rate, but it would be reasonable to assume that infected cells are more easily recognised and killed: as a consequence, the immune response would be more likely to eradicate the infection without eradicating the tumour. Furthermore, our model neglects spatial constraints or immunorefractory aspects of some tumour microenvironments that could affect the viral diffusion and the immune infiltration inside the tumour, which are well-known obstacles to the success of both virotherapy (as already discussed in Ref. \cite{morselli23}) and immunotherapy \cite{almeida22} administered alone. On the other hand, in the present work, we only take into account an increase in the immune killing rate that resembles a generic immune-boosting therapy, such as immune checkpoint inhibition. Several immunotherapies have shown their success when combined with oncolytic virotherapy \cite{shi20}, such as, but not limited to, adoptive T-cell transfer \cite{krabbe21}, CAR-T \cite{he23}, CAR-T and BiTE \cite{wing18}. It could, therefore, be interesting to analyse whether the combination of different immunotherapies could partially overcome the above-mentioned obstacles.

Another potentially interesting addition to the model is the role of hypoxia, whose interaction with oncolytic virotherapy is not straight-forward: while some viruses are able to specifically target hypoxic cells, some others are unable to effectively infect hypoxic regions due to the reduced protein translation of those cells \cite{shengguo11}. Hypoxia-driven inflammation constitutes an additional challenge when considering the interactions with the immune system.

From the mathematical point of view, it could be interesting to perform a rigorous analysis of the PDE that we have obtained. It is well-known that chemotactic models may lead to blow-up in finite time \cite{hillen09}, hence the well-posedness for long times may not be completely trivial. Another interesting future direction is the formulation of a hybrid framework that uses a deterministic model for large densities and switch to a stochastic model for very low densities. To our knowledge, this approach has been only used in nonspatial models \cite{duncan16} or in spatial models with well-defined subdomains \cite{smith18}: the use of this technique in our context does not appear straight-forward, but it could potentially complement the findings of the present work.

\newpage
\appendix
\section{Formal derivation of continuum models}
\label{app:derivation}

In this Appendix we describe how to derive the model discussed in the main text.

\subsection{Uninfected cancer cells}
Uninfected cells can first move, then reproduce or die based on the pressure value, then become infected and finally be killed by immune cells, as explained in Section \ref{sec:model}. The principle of mass balance gives the equation
\[
\begin{split}
	u_j^{n+1}&=\Bigl[\frac{\theta}{2} u_{j-1}^n+\frac{\theta}{2} u_{j+1}^n+(1-\theta) u_{j}^n\Bigr] \Bigl[1+\tau G(\rho_j^n)_+-\tau G(\rho_j^n)_-\Bigl] \\
	&\phantom{=}\times  \Bigl(1-\tau \frac{\beta}{K} i_j^n\Bigr) \Bigl(1-\tau \frac{\zeta}{K} z_j^n\Bigr),
\end{split}
\]
\rev{where $\rho_j^n\coloneqq u_j^n+i_j^n$. U}sing the algebraic relation $x_+-x_-=x$, this simplifies to
\[
\begin{split}
	u_j^{n+1}&=\Bigl[\frac{\theta}{2} u_{j-1}^n+\frac{\theta}{2} u_{j+1}^n+(1-\theta) u_{j}^n\Bigr] \Bigl[1+\tau G(\rho_j^n)\Bigl]\\
	&\phantom{=}\times\Bigl(1-\tau \frac{\beta}{K} i_j^n\Bigr) \Bigl(1-\tau \frac{\zeta}{K} z_j^n\Bigr).
\end{split}
\]
Let us define
\begin{equation*}
	\Phi\coloneqq \frac{\theta}{2}u_{j-1}^n+\frac{\theta}{2} u_{j+1}^n -\theta u_j^n=\frac{\theta}{2} \delta^2 \frac{u_{j-1}^n+u_{j+1}^n-2u_j^n}{\delta^2},
\end{equation*}
so that the previous equation becomes
\begin{align*}
	u_j^{n+1}&=(u_j^n+\Phi)\Bigl[1+\tau G(\rho_j^n)\Bigl]\Bigl(1-\tau \frac{\beta}{K} i_j^n\Bigr) \Bigl(1-\tau \frac{\zeta}{K} z_j^n\Bigr)\\
	&=u_j^n+\tau G(\rho_j^n) u_j^n-\tau \frac{\beta}{K} u_j^n i_j^n -\tau \frac{\zeta}{K} u_j^n z_j^n+\Phi \\
	&-\tau^2 G(\rho_j^n) \frac{\beta}{K} u_j^n i_j^n -\tau^2 G(\rho_j^n) \frac{\zeta}{K} u_j^n z_j^n +\tau^2\frac{\beta\zeta}{K^2} u_j^n i_j^n z_j^n +\tau^3 G(\rho_j^n)\frac{\beta\zeta}{K^2} u_j^n i_j^n z_j^n\\
	&\phantom{=}+ \tau \Phi \Bigl[ G(\rho_{j}^n)-\frac{\beta}{K} i_j^n  -\frac{\zeta}{K} z_j^n- \tau G(\rho_{j}^n)\frac{\beta}{K} i_j^n -\tau G(\rho_j^n) \frac{\zeta}{K} z_j^n +\tau\frac{\beta\zeta}{K^2} i_j^n z_j^n \\
	&\phantom{=}+\tau^2 G(\rho_j^n)\frac{\beta\zeta}{K^2} i_j^n z_j^n\Bigr]. 
\end{align*}
We now divide both sides of the previous equation by $\tau$ and rearrange the terms to get
\begin{equation}
	\label{eq:u_almostdoneIV}
	\frac{u_j^{n+1}-u_{j}^n}{\tau}=G(\rho_j^n)u_{j}^n- \frac{\beta}{K} u_j^n i_j^n - \frac{\zeta}{K} u_j^n z_j^n+\frac{1}{\tau} \Phi +H_1,
\end{equation}
and observe that every term of $H_1$ is multiplied either by $\tau$ or by $\Phi$.

Let us now assume that there is a function $u\in C^2([0,+\infty)\times\R)$ such that $u_{j}^{n}=u(t_n,x_j)=u$ (from now on we omit the arguments of functions computed at $(t_n,x_j)$); thus, we can use Taylor expansions for $u$ in time and space as follows
\begin{gather*}
	u_{j}^{n+1}=u(t_n+\tau,x_j)=u+\tau \pt u+ \bigO(\tau^2),\\
	u_{j\pm1}^{n}=u(t_n,x_j\pm\delta)=u\pm \delta \px u+ \frac{1}{2} \delta^2 \pxx u+\bigO(\delta^3).
\end{gather*}
This implies that 
\[
\Phi=\frac{\theta}{2} \delta^2 \pxx u+\bigO(\delta^3),
\]
and thus $H_1= \bigO(\tau)+\bigO(\delta^2)$. Eq. \eqref{eq:u_almostdoneIV} then becomes
\[
\pt u+\bigO(\tau^2)=\theta\frac{\delta^2}{2\tau} \pxx u +G(\rho) u-\frac{\beta}{K} ui-\frac{\zeta}{K} uz+\bigO(\tau)+\bigO(\delta^2).
\]
Letting $\tau,\delta\to 0$ in such a way that $\frac{\delta^2}{2\tau}\to \tilde{D}$, we obtain
\[
\pt u=\theta \tilde{D}\pxx u+G(\rho)u-\frac{\beta}{K} ui -\frac{\zeta}{K} uz.
\]

\subsection{Infected cancer  cells}

Infected cells can first move, then die, as explained in Section \ref{sec:model}. Also, uninfected cells may be infected. The principle of mass balance gives the equation
\begin{align*}
	i_j^{n+1}&=\Bigl[\frac{\theta}{2}i_{j-1}^n+\frac{\theta}{2}i_{j+1}^n+(1-\theta) i_{j}^n\Bigr](1-\tau q) \Bigl(1-\tau \frac{\zeta}{K} z_j^n\Bigr) \\
	&\phantom{=}+\tau \frac{\beta}{K} i_j^n (1+\tau G(\rho_j^n)) \Bigl(1-\tau \frac{\zeta}{K} z_j^n\Bigr)( u_{j}^n +\Phi),
\end{align*}
which simplifies to
\begin{align*}
	i_j^{n+1}&=(1-\tau q)\Bigl(1-\tau \frac{\zeta}{K} z_j^n\Bigr) (i_j^n+\Psi)+\tau \frac{\beta}{K} u_j^n i_j^n+\tau H_2\\
	&=i_j^n-\tau q i_j^n -\tau \frac{\zeta}{K} i_j^nz_j^n+\Psi +\tau H_3\\
	&\phantom{=}+\tau \frac{\beta}{K} u_j^n i_j^n+\tau H_2, 
\end{align*}
where
\[
\Psi\coloneqq \frac{\theta}{2}i_{j-1}^n+\frac{\theta}{2} i_{j+1}^n -\theta i_j^n=\frac{\theta}{2}\delta^2 \frac{i_{j-1}^n+i_{j+1}^n-2i_j^n}{\delta^2}.
\]
and
\begin{gather*}
	H_2\coloneqq \tau  G(\rho_j^n) \frac{\beta}{K} u_j^n i_j^n -\tau \frac{\beta\zeta}{K^2} u_j^n i_j^n z_j^n -\tau^2 G(\rho_j^n) \frac{\beta\zeta}{K^2} u_j^n i_j^n z_j^n\\
	+\frac{\beta}{K} i_j^n (1+\tau G(\rho_{j}^n)) \Bigl(1-\tau \frac{\zeta}{K} z_j^n\Bigr) \Phi, 
	\\
	H_3\coloneqq -q\Psi -  \frac{\zeta}{K} z_j^n \Psi +\tau q \frac{\zeta}{K} i_j^n z_j^n +\tau q \frac{\zeta}{K} z_j^n \Psi.
\end{gather*}
Let us observe that every term of $H_2$ and $H_3$ is multiplied either by $\tau$, $\Phi$ or $\Psi$. Dividing both sides by $\tau$ and rearranging the terms, we get 
\begin{equation}
	\label{eq:i_almostdoneIV}
	\frac{i_j^{n+1}-i_j^n}{\tau}=\frac{1}{\tau}\Psi-qi_j^n+\frac{\beta}{K} u_j^n i_j^n-\frac{\zeta}{K} i_j^nz_j^n+H_2+H_3.
\end{equation}

Let us now assume that there is a function $i\in C^2([0,+\infty)\times\R)$ such that $i_{j}^{n}=i(t_n,x_j)=i$, so that
\begin{gather*}
	i_{j}^{n+1}=i(t_n+\tau,x_j)=i+\tau \pt i+ \bigO(\tau^2),\\
	i_{j\pm1}^{n}=i(t_n,x_j\pm\delta)=i\pm \delta \px i+ \frac{1}{2} \delta^2 \pxx i+\bigO(\delta^3).
\end{gather*}
This implies that 
\[
\Psi=\frac{\theta}{2} \delta^2 \pxx i+\bigO(\delta^3),
\]
and thus $H_2+H_3= \bigO(\tau)+\bigO(\delta^2)$. Eq. \eqref{eq:i_almostdoneIV} then becomes
\[
\pt i+\bigO(\tau^2)=\theta\frac{\delta^2}{2\tau} \pxx i +\frac{\beta}{K} ui-\frac{\zeta}{K} iz-qi+\bigO(\tau)+\bigO(\delta^2).
\]
Letting $\tau,\delta\to 0$ in such a way that $\frac{\delta^2}{2\tau}\to \tilde{D}$ we obtain 
\[
\pt i=\theta \tilde{D}\pxx i+\frac{\beta}{K} ui-\frac{\zeta}{K} iz-qi.
\]

\subsection{Chemoattractant}
The chemoattractant is produced by cancer cells, decays at a constant rate and diffuses, as explained in Section \ref{sec:model}; its dynamics are described by Eq. \eqref{eq:phi_discrete}, which can be written as 
\begin{equation}
	\frac{\phi_j^{n+1}-\phi_j^n}{\tau}= D_\phi \frac{\phi_{j+1}^n+\phi_{j-1}^n-2\phi_j^n}{\delta^2} + (\alpha_\phi i_j^n+\gamma_\phi u_j^n) \rev{(\phi^*-\phi_j^n)}- q_\phi \phi_j^n
\end{equation}
Let us now assume that there is a function $\phi\in C^2([0,+\infty)\times\R)$ such that $\phi_{j}^{n}=\phi(t_n,x_j)$ $=\phi$. Then clearly the discrete diffusion term converges to the second derivative of $\phi$, hence for $\tau,\delta\to 0$ we obtain
\[
\pt \phi= D_\phi \pxx\phi+ (\alpha_\phi i+\gamma_\phi u) \rev{(\phi^*-\phi)}- q_\phi \phi.
\]

\subsection{Immune cells}

Immune cells can first move, then die, as explained in Section \ref{sec:model}. Also, new immune cells may enter the domain. The principle of mass balance gives the equation
\[
z_j^{n+1}=\Bigl[F_{j-1\to j}^n\, z_{j-1}^n+F_{j+1\to j}^n\, z_{j+1}^n+(1-F_{j\to j-1}^n-F_{j\to j+1}^n) z_{j}^n\Bigr] (1-\tau q_z)+ \tau S_j^n,
\]
with
\[
F_{j\to j\pm 1}^n\coloneqq\frac{\theta_z}{2} +\underbrace{\nu\frac{(\phi_{j\pm 1}^n-\phi_j^n)_+}{2\phi^*}}_{\eqqcolon \tilde{F}_{j\to j\pm 1}^n}, \qquad S_j^n=\biggl(S_0+\alpha_z  \sum_h I_h^n\biggr) \mathbbm{1}_\omega (x_j).
\]
Observe that the source term is no longer multiplied by $\delta$, since we are considering the cell density. The previous equation can be written as 
\[
z_j^{n+1}=(z_j^n+\Xi)(1-\tau q_z)+ \tau S_j^n =z_j^n+\Xi-\tau q_z z_j^n-\tau q_z \Xi +\tau S_j^n,
\]
where
\begin{gather*}
	\Xi\coloneqq \Xi_1+\Xi_2, \\
	\Xi_1\coloneqq \theta_z \frac{z_{j-1}^n+z_{j+1}^n-2z_j^n}{2}, \\
	\Xi_2\coloneqq -(\tilde{F}_{j\to j-1}^n+\tilde{F}_{j\to j+1}^n)z_j^n+\tilde{F}_{j-1\to j}^n z_{j-1}^n+\tilde{F}_{j+1\to j}^n z_{j+1}^n,
\end{gather*}
Dividing both sides by $\tau$ and rearranging the terms, we get 
\begin{equation}
	\label{eq:z_almostdoneIV}
	\frac{z_j^{n+1}-z_j^n}{\tau}= \frac{\Xi}{\tau} -q_z z_j^n +S_j^n +q_z \Xi.
\end{equation}

Let us now assume that there is a function $z\in C^2([0,+\infty)\times\R)$ such that $z_{j}^{n}=z(t_n,x_j)=z$, so that
\begin{gather*}
	z_{j}^{n+1}=z(t_n+\tau,x_j)=z+\tau \pt z+ \bigO(\tau^2),\\
	z_{j\pm1}^{n}=z(t_n,x_j\pm\delta)=z\pm \delta \px z+ \frac{1}{2} \delta^2 \pxx z+\bigO(\delta^3),
\end{gather*}
This implies that
\[
\Xi_1=\frac{\theta_z}{2} \delta^2 \pxx z+\bigO(\delta^3).
\]
Furthermore, the assumptions on $\phi$ imply that 
\[
\tilde{F}_{j\to j\pm 1}^n=\frac{\nu}{2\phi^*}\Bigl(\pm\delta\px \phi+\frac{1}{2}\delta^2\pxx \phi+\bigO(\delta^3)\Bigr)_+=\bigO(\delta).
\]
and we can easily conclude that $\Xi=\bigO(\delta)$. We then use the Taylor expansion of $z$ in the definition of $\Xi_2$ to get
\[
\begin{split}
	\Xi_2	&=-(\tilde{F}_{j\to j-1}^n+\tilde{F}_{j\to j+1}^n)z+\tilde{F}_{j-1\to j}^n\Bigl(z-\delta\px z+\frac{1}{2}\delta^2\pxx z+\bigO(\delta^3)\Bigr)\\
	&\phantom{=}+\tilde{F}_{j+1\to j}^n \Bigl(z+\delta\px z+\frac{1}{2}\delta^2\pxx z+\bigO(\delta^3)\Bigr)\\
	&=(\tilde{F}_{j-1\to j}^n-\tilde{F}_{j\to j-1}^n+\tilde{F}_{j+1\to j}^n-\tilde{F}_{j\to j+1}^n)z+\delta(-\tilde{F}_{j-1\to j}^n+\tilde{F}_{j+1\to j}^n)\px z\\
	&\phantom{=}+\frac{1}{2}\delta^2(\tilde{F}_{j-1\to j}^n+\tilde{F}_{j+1\to j}^n)\pxx z +\bigO(\delta^3).
\end{split}
\]
Now, let us observe that
\begin{align*}
	\tilde{F}_{j\pm1\to j}^n-\tilde{F}_{j\to j\pm1}^n&=\frac{\nu}{2\phi^*}[(\phi_j^n-\phi_{j\pm 1}^n)_+ - (\phi_{j\pm 1}^n-\phi_{j}^n)_+]\\
	&=\frac{\nu}{2\phi^*}(\phi_{j}^n-\phi_{j\pm 1}^n)=
	\frac{\nu}{2\phi^*}\Bigl(\mp\delta\px \phi-\frac{1}{2}\delta^2\pxx \phi+\bigO(\delta^3)\Bigr),
\end{align*}
using the relation $x_+-(-x)_+=x_+-x_-=x$. We therefore have
\[
\Xi_2=\frac{\nu}{2\phi^*}\Bigl\{-\delta^2\pxx \phi z+\delta[-(\delta\px\phi+\bigO(\delta^2))_+ +(-\delta\px\phi+\bigO(\delta^2))_+]\px z+\bigO(\delta^3)\Bigr\}.
\]
Finally, Eq. \eqref{eq:z_almostdoneIV} becomes
\begin{align*}
	\pt z+ \bigO(\tau^2)&=\theta_z\frac{\delta^2}{2\tau} \pxx z+\bigO(\frac{\delta^3}{\tau})+\frac{\nu}{\phi^*}\ \frac{\delta^2}{2\tau}\Bigl\{-\pxx \phi z\\
	&\phantom{=}+[-(\px\phi+\bigO(\delta))_++(-\px\phi+\bigO(\delta))_+]\px z+\bigO(\delta)\Bigr\}-q_z z+S_j^n+\bigO(\delta).
\end{align*}
Let us now observe that 
\[
\sum_h I_h^n=\sum_h i_h^n \,\delta =\sum_h i(t_n,x_h)\delta \xrightarrow{\delta\to 0} \int_\Omega i(t_n,y) \di y,
\]
which means that
\[
S_j^n= \biggl(S_0+\alpha_z  \sum_h I_h^n\biggr) \mathbbm{1}_\omega (x_j) \xrightarrow{\delta\to 0} \biggl(S_0+\alpha_z  \int_\Omega i(t_n,y) \di y\biggr) \mathbbm{1}_\omega (x_j)\eqqcolon S(t_n,x_j).
\]
Letting $\tau,\delta\to 0$ in such a way that $\frac{\delta^2}{2\tau}\to \tilde{D}$ we arrive at the final result:
\[
\begin{split}
	\pt z&=\theta_z \tilde{D}\pxx z+\frac{\nu \tilde{D}}{\phi^*}\{-\pxx \phi z+[-(\px\phi)_++(-\px\phi)_+]\px z\}-q_z z+S\\
	&=\theta_z \tilde{D}\pxx z+\frac{\nu \tilde{D}}{\phi^*}(-\pxx \phi z-\px\phi\px z)-q_z z+S\\
	&=\theta_z \tilde{D}\pxx z-\frac{\nu \tilde{D}}{\phi^*}\px (z\px\phi)-q_z z+S.
\end{split}
\]

\section{Oncolytic viral infection in absence of the immune response: main results from Ref. \cite{morselli23}}
\label{subsec:bmb}

As explained in the main text, the models presented in this article are an extension of some of the models introduced in Ref. \cite{morselli23}, i.e., Eq. \eqref{eq:l} and its underlying stochastic counterpart. For the sake of completeness, we here summarise the main takeaways, which partially apply also to the current models. As the PDE model agrees well with the agent-based simulations, we can use our knowledge of the continuous model to better understand the outcome of the therapy in different parameter regimes. 

The worst possible case is the situation in which the infection ceases after a finite time and uninfected cells grow at carrying capacity: this corresponds to parameter values such that $\beta<q$, which do not allow the equilibrium $(u^*,i^*)$ to be positive.

Let us now focus on the case $\beta>q$. Initially, the tumour centre is rapidly infected while uninfected cells at the outer edge grow until carrying capacity. The uninfected front advances at speed  $2\sqrt{D p}$ and once carrying capacity is reached, the infected front stabilises at  $2\sqrt{D(\beta-q)}$. Both speed values are in line with theoretical results: indeed, the value $2\sqrt{D p}$ is the well-known wave speed of the Fisher-KPP equation \cite{fisher37,kolmogorov37}; the value $2\sqrt{D(\beta-q)}$ can be easily obtained using standard linearisation techniques and is in line to the rigorous analytical result proved for a slightly different system in Ref. \cite{dunbar84}. While the tumour and the infections expand, cell densities at the centre of the tumour converge with damped oscillations to the equilibrium of the corresponding ODE, given by Eq.~\eqref{eq:eq_l}.

If $\beta<q+p$, the speed of the infection is smaller than the speed of the uninfected wave and so the outer region of the tumour is completely unaffected by the therapy: this is a failure that has no analogue in the spatially homogeneous ODE and can be only predicted through the spatial model. On the other hand, whenever $\beta>q+p$, the infection eventually reaches the front of the wave of uninfected cells. The final peak of the uninfected cells is approximately 
\begin{equation}
	\label{eq:ubar}
	\bar{u}\coloneqq \Bigl( \frac{q}{\beta}+\frac{p}{\beta} \Bigr) K
\end{equation}
which allows the infected front to move at speed $2\sqrt{D p}$. In this last case, we can consider the therapy to be at least partially successful; some variations of the parameter values then allow for the improvement of therapy achievements. Overall, the efficacy of the infection increases as $\beta$ increases and $q$ decreases, as it is pointed out in the previous suggestion related to the speed of the waves also suggested by the equilibrium values given in Eq. \eqref{eq:eq_l}.

\section{Details of numerical simulations}
\label{app:num}

\paragraph{Parameter values}

The majority of the parameters of the model has been estimated from the empirical literature, while a few others are specific of our formulation of the model and have been set to reasonable values in order to reproduce plausible dynamics. The parameters $p, D, K$ and $\beta$ assume the values listed in Table \ref{tab:parameters}, which are the same used in Ref. \cite{morselli23}; for the sake of brevity, we omit further comments on them and report only the references in the aforementioned table. On the other hand, the basic death rate of infected cells $q$ has been decreased to $8.34\times 10^{-3}\;$h$^{-1}$, which is one-fifth of the value used in Ref. \cite{morselli23}. This is due to the fact that, in the current model, $q$ does not take into account the death of infected cells due to immune killing, which is considered separately.

The diffusion coefficient of the chemoattractant $D_\phi$ has been taken equal to $3.33\times 10^{-2}\;$mm$^2$/h following Ref. \cite{matzavinos04}. The saturation density of the chemoattractant $\phi^*$ and the secretion rates of chemoattractant by infected cells $\alpha_\phi$  have been adapted from the values reported in Ref. \cite{jenner22}, which fit the data of IFN $\gamma$ taken from Ref. \cite{gao14}. The value of $\phi^*$ has been obtained by rescaling the estimate of Ref. \cite{jenner22} to our two-dimensional setting, yielding a value of $2.92\;\mu$g/mm$^2$. The secretion rate reported in Ref. \cite{jenner22} refers to a single CD4$^+$T and these cells are assumed to be stimulated by infected tumour cells; we have adapted their value to our setting by dividing it by the carrying capacity $K$, obtaining the value $\alpha_\phi=2.50\times 10^{-4}\;$mm$^2$/(h$\cdot$cells). Since we are assuming that immune cells are much less stimulated by uninfected tumour cells, we have set the secretion rate of chemoattractant by uninfected cells $\gamma_\phi$ to $5.00\times 10^{-6}\;$mm$^2$/(h$\cdot$cells), which still allows obtaining a considerable reduction of the tumour load when the infection stimulates the immune system. The decay of chemoattractant $q_\phi$ has been taken equal to $8.3\times 10^{-2}\;$h$^{-1}$, as in Ref. \cite{cooper14}.

While it is clear from experimental results that the speed of an immune cell is around $1.08\;$mm/h \cite{textor11}, the estimate of the diffusion and chemotactic coefficients $D_z$ and $\chi$ from this consideration constitute a particular challenge. The diffusion coefficient of immune cells has been set to $D_z=4.20\times 10^{-3}\;$mm$^2$/h, as in Ref. \cite{almeida22}, noting that similar values are used elsewhere in the literature (such as in Ref. \cite{atsou20}). We remark that, following the approach of Refs. \cite{hillenswan16,othmer02}, one could estimate a value in the same order of magnitude relying on reasonable biological assumptions (although the precise quantities needed are hard to estimate). We performed several simulations of our agent-based model to conclude that for $\chi=1.65\;$mm$^2$/h immune cells move toward a gradient of chemoattractant similar to the one present in our simulations (i.e., the stationary profile of the chemokines for the initial condition of our simulations) with an average speed of approximately $0.6\;$mm/h. We decrease this value to $\chi=0.165\;$mm/h$^2$ to avoid an excessive concentration of immune cells; the resulting average speed is around $0.06\;$mm/h, which is plausible considering that immune cells face many physical obstacles in penetrating the tumour microenvironment.

The death rate of immune cells $q_z$ has been taken equal to $7.5\times10^{-3}\;$h$^{-1}$, which is the value used in Ref. \cite{hao2014mathematical} for T4 cells. The base inflow rate of immune cells $S_0$ has been set to $5.00\times 10^{-2}$ cells/(mm$^2\cdot$h) in order to have a density of immune cells inside the tumour coherent with experimental observations \cite{chatzopoulos20,yasuda11}. The additional inflow rate of immune cells due to the infection $\alpha_z$ is one of the main peculiarities of our model and summarises several biological processes; hence, it is hard to find meaningful estimates in the literature. We have set it to $3.75\times 10^{-5}\;$(mm$^2\cdot$h)$^{-1}$, which allows us to have an immune cell density comparable with the aforementioned experimental references. The same problem arises with the immune killing rate of cancer cells $\zeta$: since the model is very sensitive to this parameter, we compare the differences between setting it to $0.50\;$h$^{-1}$ and  $5.00\;$h$^{-1}$, corresponding respectively to weak and enhanced immune responses.

\paragraph{Numerical simulations for the discrete models}

We used a temporal step $\tau=0.02\;$h and a spatial step of $\delta=0.1\;$mm, as already done in Ref. \cite{morselli23}. All simulations have been performed in \textsc{Matlab 2021b}.

At every iteration, the cell numbers and the chemoattractant density are updated according to the rules described in Section \ref{sec:model}. We first consider movement, then reproductions and deaths of all cell populations, the inflow of immune cells, infections and finally chemoattractant dynamics. Since we only need to keep track of the collective fate of cells in the same lattice point, we again used the built-in \textsc{Matlab} functions \texttt{binornd} and \texttt{mnrnd}. Zero-flux boundary conditions for cell populations are implemented by not allowing cells at the boundary to leave the domain. The density of the chemoattractant is updated through the two-dimensional analogue of Eq. \eqref{eq:phi_discrete}; Neumann boundary conditions are then implemented by considering additional grid points outside the domain, with the same density value as the boundary points of the grid.

The one-dimensional plots in Fig. \ref{fig:basics}a-b are obtained by averaging ten simulation. The cell sums of Fig. \ref{fig:comparison} are obtained by averaging five simulations (although the cell sum obtained from a single simulation does not show any significant difference). The ICP and TCP plots in Figs. \ref{fig:eradication_infection} and \ref{fig:violin}, as well as the violin plots of Fig. \ref{fig:violin}, show the result of one hundred simulations. All the two-dimensional plots show a single simulation. We remark that, in all cases, we performed at least five simulations and did not observe any relevant qualitative difference with respect to the result shown; the only exception is Fig. \ref{fig:violin} and the electronic supplementary material  S6, as explained in the main text. 

In order to allow reproducibility, a random seed has been set at the beginning of each new simulation. In the figures representing a single simulation, only the one with random seed equal to 1 is shown (with the exception of the electronic supplementary material  S6  in which it has been set to 4).

\paragraph{Numerical simulations for the continuum models}
The system of equations \eqref{eq:pderad} has been solved with a finite difference scheme explicit in time, using the discretisations $\Delta t=10^{-4}$, $\Delta x=0.01$; such a low space step allows to appropriately describe the high peaks of immune cells of some simulations. The only exceptions are the simulations of Fig. \ref{fig:bif_p}, which are run for a very long time in a bigger domain: there, the discretisation for space is $\Delta x=0.02$, which guarantees stability at late times without the need to decrease the time step. We used a forward upwind scheme for the chemotactic term in the equation for immune cell density, following Ref. \cite{leveque07}; this is a common strategy to deal with this kind of equations \cite{almeida22,bubba20}. \rev{The integrals are computed through the built-in \textsc{Matlab} function \texttt{trapz}, which is based on a linear interpolation of functions.} We also use again the threshold $\frac{1}{\delta^2}$ to identify the wavefront of infected cells in the solution of the PDE; this allows us to be consistent with the representation of the agent-based model.

\section*{Supplementary information}
	
The supplementary material is available at the link \url{https://doi.org/10.5281/zenodo.18340945}.

\bibliographystyle{siam}
\bibliography{biblio}
	
\end{document}